\documentclass[a4paper,fleqn,usenatbib]{mnras}

\usepackage{multirow}

\usepackage[T1]{fontenc}
\usepackage{ae,aecompl}
\usepackage{graphicx}	
\usepackage{amsmath}	
\usepackage{amssymb}	
\usepackage{bm} 

\newcommand{\lsim}{\ \raise -2.truept\hbox{\rlap{\hbox{$\sim$}}\raise
    5.truept\hbox{$<$}\ }} \newcommand{\gsim}{\ \raise
  -2.truept\hbox{\rlap{\hbox{$\sim$}}\raise 5.truept\hbox{$>$}\ }}

\title[The impact of He abundance on the Cepheid PLRs and PWRs
 and the Distance Ladder]{On the impact of Helium abundance on the
  Cepheid Period-Luminosity and Wesenheit relations and the Distance
  Ladder}

\author[R. Carini et al.]{
R. Carini,$^{1}$\thanks{E-mail: roberta.carini@oa-roma.inaf.it}
E. Brocato,$^{1}$
G. Raimondo,$^{2}$
and M. Marconi$^{3}$
\\
$^{1}$INAF-Osservatorio Astronomico di Roma, via Frascati 33, 00078 Monte Porzio Catone, Italy\\
$^{2}$INAF-Osservatorio Astronomico di Teramo, Mentore Maggini s.n.c., 64100 Teramo, Italy\\
$^{3}$INAF-Osservatorio Astronomico di Capodimonte, Salita Moiariello 16, I-80131 Napoli, Italy
}

\date{Accepted XXX. Received YYY; in original form ZZZ}

\pubyear{2016}

\begin{document}
\label{firstpage}
\pagerange{\pageref{firstpage}--\pageref{lastpage}}
\maketitle

\begin{abstract}
  This work analyses the effect of the Helium content on synthetic
  Period-Luminosity Relations (PLRs) and Period-Wesenheit Relations
  (PWRs) of Cepheids and the systematic uncertainties on the derived
  distances that a hidden population of He-enhanced Cepheids may
  generate. We use new stellar and pulsation models to build a
  homogeneous and consistent framework to derive the Cepheid
  features. The Cepheid populations expected in synthetic
  colour-magnitude diagrams of young stellar systems (from 20\,Myr to
  250\,Myr) are computed in several photometric bands for $Y=0.25$ and
  $Y=0.35$, at a fixed metallicity ($Z=0.008$). The PLRs appear to be
  very similar in the two cases, with negligible effects (few \%) on
  distances, while PWRs differ somewhat, with systematic uncertainties
  in deriving distances as high as $\sim$\,7\% at $\log
  P<1.5$. Statistical effects due to the number of variables used to
  determine the relations contribute to a distance systematic error of
  the order of few percent, with values decreasing from optical to
  near-infrared bands.  The empirical PWRs derived from
  multi-wavelength datasets for the Large Magellanic Cloud (LMC) is in
  a very good agreement with our theoretical PWRs obtained with a
  standard He content, supporting the evidence that LMC Cepheids do
  not show any He effect.

\end{abstract}


\section*{Introduction}
\ 
Through the period-luminosity relations (PLRs), classical Cepheids are
fundamental standard candles for the calibration of secondary distance
indicators (as Supernovae Ia, surface brightness fluctuations,
etc...), for the determination of the cosmological distances and the
Hubble constant \citep{freedman01,saha}.
In spite of this, the general question of the dependence of the
Cepheid properties and consequently their PLRs on the chemical
composition, discussed by many authors over the years, still lacks of
firm conclusions, and the size and even the sign of the effects are
still disputed \citep[see e.g.][]{sasselov, kennicutt, bms99, fio02,
  storm04,m05,romaniello08,freedman10,ef14}.
In particular, the effect of He abundance on Cepheid pulsation
properties has been theoretically investigated by \cite {b00},
\cite{fio02} and \cite{m05} for $Z\ge{0.02}$. At fixed metallicity, an
increase in the He content produces a shift towards higher effective
temperature. If both the metallicity and the He abundances increase,
as expected, the two effects on the instability-strip topology tend to
compensate each other. Moreover, at fixed He to metal enrichment ratio
($\Delta{Y}/{\Delta{Z}}$), as the metallicity increases from $Z=0.03$
to $Z=0.04$ the instability strip tends to narrow. As a result of the
complex interplay between He and metal variations, in the case of
metal-rich ($Z \geq 0.01$) Cepheid samples, the predicted correction
for metallicity to the LMC-based PLR was found to be dependent on the
assumed $\Delta{Y}/{\Delta{Z}}$ (see \citealt{fio02} and \citealt{m05}
for details) and to show an opposite sign when comparing MW and MC
Cepheids, in agreement with \cite{romaniello08}.  To reduce the
influence of the instability-strip topology and of the contribution of
metals, the period-Wesenheit relations (PWRs, \citealt{madore82}) are
often adopted. These relations have the additional advantage of widely
removing the reddening effect, and of significantly reducing the
dispersion of visual magnitudes at a given period. In fact, the
dispersion of the PWRs for the LMC Cepheids is $\sim$2$\div$3 times
smaller compared to the optical PLRs
\citep{tanvir,udalski,fouque,sosz08,ngeow09}. This reduces the random
errors in the evaluation of distances. It is worth noting that the
PWRs rely on the assumed extinction law, and any deviation of the
extinction coefficients could introduce systematic errors on the
inferred distances, with major impact in star forming regions, like 30
Doradus (see e.g. \citealt{demarchi2016}). The effect of variations in
the He content on the PWRs has been predicted by \citet{carini} at the
LMC typical metal content ($Z \simeq 0.008$). In that paper, we showed
that, at fixed metallicity, the instability strip of fundamental
pulsators becomes hotter when the He increases from the standard value
$Y=0.25$ to $Y=0.35$. At fixed mass, Cepheids with higher He abundance
pulsate with longer periods and, consequently, the PWRs have different
slopes with respect to those of standard relations, as shown in
Fig.~10 and Table~5 in \citet{carini}.  In this paper, we analyse the
effect of He content on synthetic PLRs and PWRs and on the inferred
Cepheid-based distance determinations, by relying on stellar
pulsation, stellar evolution and population synthesis models. We
explore the role He-enhanced Cepheids might have if present in a
sizable fraction. In particular, we will verify whether and how the
PLRs and PWRs may be affected by the presence of Cepheids with
non-standard He contents, through comparisons with samples of Cepheids
observed in the LMC.  This has relevant implications in view of the
present efforts to reach the 1\% level to properly confront the local
value of H$_{0}$ with the microwave background measurements.

Moreover, it is now well known that multiple stellar populations
(MSPs) are present in the globular clusters (GCs) of the Milky Way
(MW, see the review by \citealt{piotto09}), where the chemistry of
stars belonging to the second generation(s) is altered with respect to
the abundances of the original gas from which the cluster has been
formed \citep{grat04,carretta09,grat2012}.  Some studies suggested
that the newly formed stars should be He-enhanced (up to $Y\sim0.40$,
e.g. \citealt{piotto07,grat2010,marino2014}), while their metallicity
appears nearly constant (the differences are less than 0.05 dex,
\citealt{car09b}).  Such an evidence seems to favour the hypothesis
that the progenitors of the GCs with MPs should be very massive
($M>10^5$ M$_\odot$), with second generation (SG) stars formed during
the first $\sim$ 150 million years of the cluster's life
\citep{dercole2008,decressin,demink,bastian2}.  If a similar formation
history is assumed, one expects signatures of second generation stars
in (relatively) young massive clusters.  For instance, Cepheids with
different chemical content (with respect to the bulk of stars) might
exist as well, even if to date they have not been detected.
He-enhanced variables in observed samples of Cepheids could, in
principle, alter the slope and zero point of PLRs and PWRs, with
respect to those derived from He-standard variables.  Unfortunately,
observations available up to now do not support or exclude in a
conclusive way the presence of SG He-enhanced stars in young stellar
systems.  Some intermediate-age ($\sim$ 1-2 Gyr) clusters of the Large
Magellanic Cloud (LMC) show distinct main-sequence turn-offs, as well
as few younger clusters \citep{milone2016}. This occurrence has been
interpreted as a sign of (at least) two main episodes of star
formation, although alternative explanations have been invoked
\citep[different rotation rates, binaries, etc ..., e.g.][]{milone09,milone1866}.
The picture is still uncompleted and deserve more investigations using
different approaches.  In this paper we make a theoretical
investigation to explore the impact of multiple populations with
different He content on the observed Cepheids properties of a stellar
system.

The paper is arranged as follow: Section~\ref{sec:model} describes the
stellar evolution and pulsation models, and introduces the adopted
stellar population synthesis models. Our theoretical PL and PW
relations for the two values of initial He are compared with the most
updated empirical ones derived from multi-wavelength datasets for the
LMC in Section~\ref{sec:observation}.
In  Section~\ref{sec:ogle}  our theoretical results are compared to  Cepheids in  LMC adopting OGLE III \citep{udalski08,sosz08} and VMC \citep{cioni11,ripepi12} releases.
In Section~\ref{sec:popeffects} we evaluate the stochastic effects on
deriving distances when only a small sample ($\sim 50$) of Cepheids is
available. A brief discussion closes the paper.

\begin{figure*}
\center
\includegraphics[trim= 1cm 1cm 1cm 1cm, clip=true,width=.9\columnwidth]{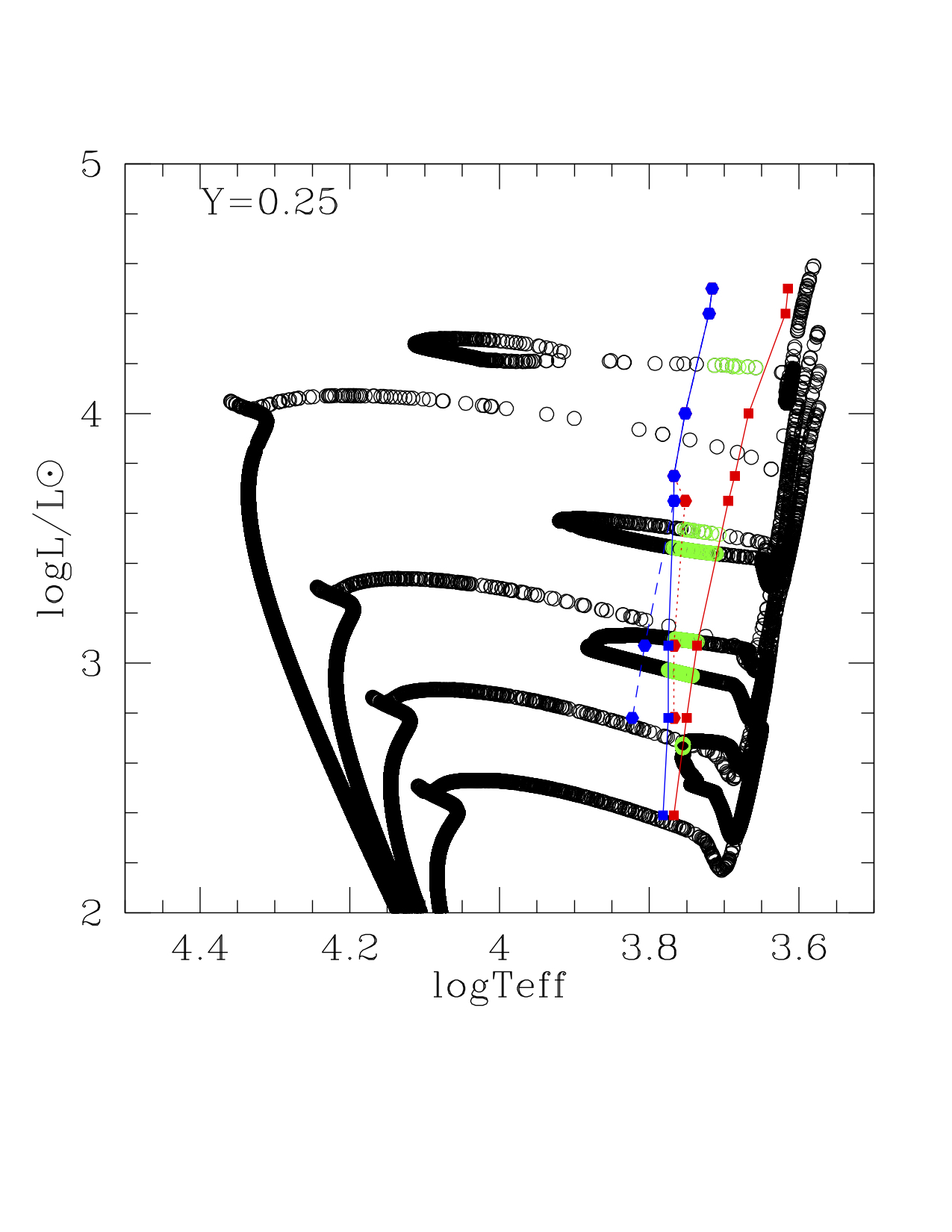}
\includegraphics[trim= 1cm 1cm 1cm 1cm, clip=true,width=.9\columnwidth]{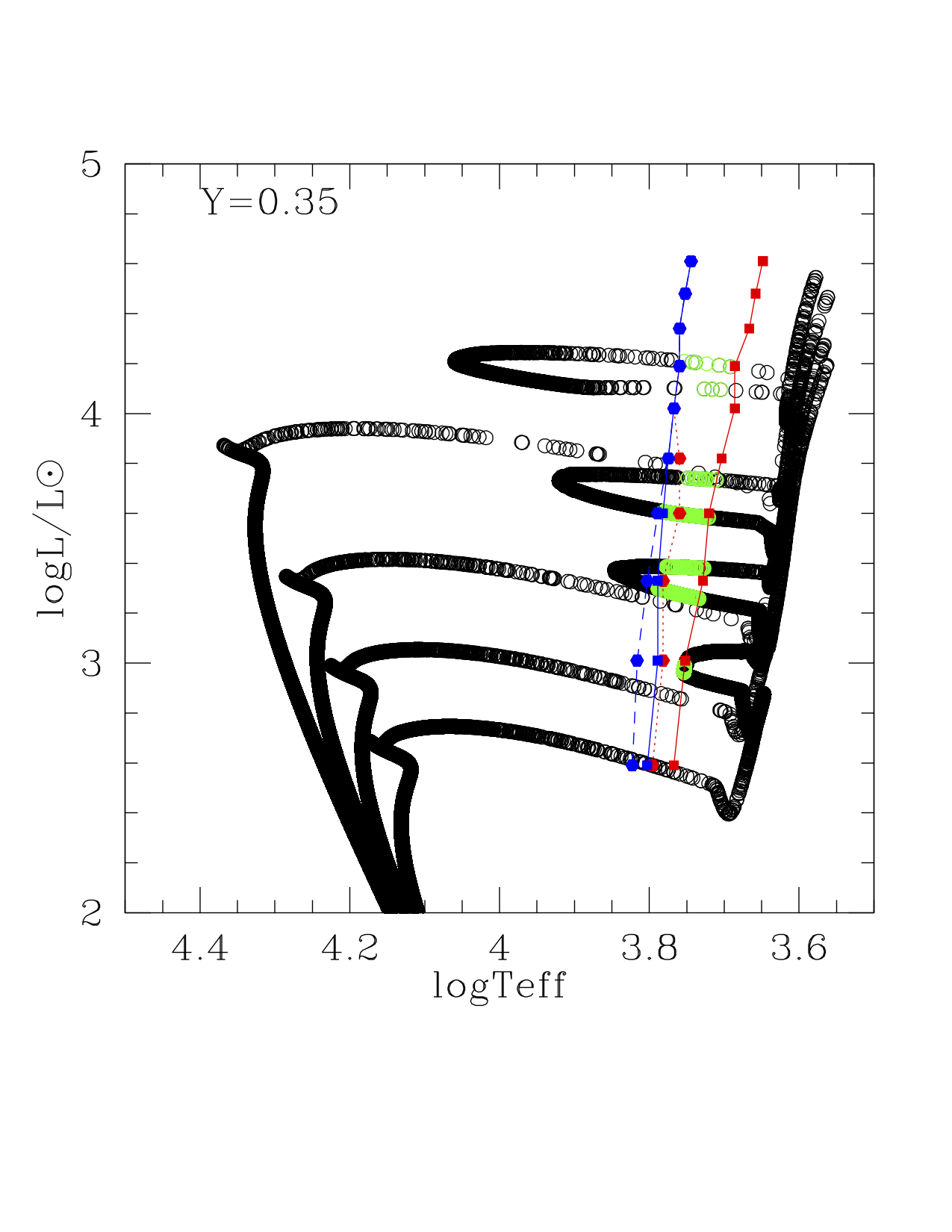}
\vspace{-1cm}
\caption{HR diagram of stars populations with $t=30$, $80$, $150$ and
  $254$ Myr for $Y=0.25$ (left panel) and $t=30$, $60$, $100$ and
  $156$ Myr for $Y=0.35$ (right panel). The FRE (solid red line), FORE
  (dotted red line), FBE (solid blue line), and the FOBE (short dashed
  blue line) are also shown. In these plots the photometric
  uncertainties are neglected. (For a colour version of the figure see
  the electronic version of the paper).}
\label{fig1}
\end{figure*}

\section{Model computations}
\label{sec:model}

\subsection{Stellar evolution and pulsation models}

Stellar evolution models have been computed using the ATON
evolutionary code \citep{ventura98} and published in
\citet{carini}. Here, we briefly recall the main assumptions of the
code. The convective instability is described by means of the Full
Spectrum of Turbulence (FST) model developed by \citet{canuto}; the
mass-loss rate in the ATON case is determined via the Bl\"ocker's
formulation \citep{blocker}:

\begin{displaymath}
\dot{M}=4.83 \times 10^{-22}\eta_RM^{-3.1}L^{3.7}R,
\end{displaymath}
where $M, L$, and $R$ are denoted in solar units, $\eta_R$ is a free
parameter. We used $\eta_R =0.02$.

The mixing of chemicals and nuclear burning is coupled with a
diffusive approach, according to the scheme suggested by
\citet{cloutman}.  The overshoot of convective eddies into radiatively
stable regions is described by an exponential decay of the velocity
beyond the formal border found via the Schwarzschild criterion.  The
extent of the overshoot is given by the free parameter $\xi$, in this
study we put $\xi=0.02$, in agreement with the calibration given in
\citet{ventura98}. No overshooting is used for the evolutionary phases
following the core He-burning phase.

The evolutionary models used in the present work have metallicity
$Z=0.008$, initial He $Y=0.25$ and $Y=0.35$; the mixture is
$\alpha$-enhanced, with [$\alpha$/Fe]$=+0.2$, with the reference solar
mixture taken from \citet{gs98}. The star evolution with both He
contents is followed from the main sequence (MS) to the beginning of
the phase of thermal pulses; the range in mass is from 0.4\,M$_\odot$
to 12\,M$_\odot$.

The sets of pulsation models computed for the above chemical
compositions in \citet{carini} have been extended with the computation
of a finer grid of stellar masses with the same nonlinear pulsation
code (see \citealt{m05,m10} and references therein). The adopted
physical and numerical assumptions are the same as in \cite{m05} ( see
also \citealt{bono98}; \citealt{bms99} for details). Here, we only
remind that the adopted nonlinear hydrodynamical code includes a
nonlocal time-dependent treatment of convection
\citep{stel82,bms99}. For each adopted stellar mass ranging from
3\,M$_\odot$ to 12\,M$_\odot$, the assumed luminosity level is based
on a canonical mass-luminosity relation \citep{b00}. A wide range of
effective temperatures is explored with a step of 100\,K in order to
investigate the pulsation stability in both the fundamental (F) and
the first overtone (FO) mode for each mass, luminosity and chemical
composition.  More details can be found in \cite{carini}.

\begin{table}
\centering
\caption{Mean value of  the mass, luminosity and temperature of the stars at the  termination of the Main Sequence luminosity function for each age.}
\label{to}
\begin{tabular}{cccc}
\hline
t (Myr) & $M/M_{\odot}$ & $\log L/L_{\odot} $ & $\log Teff $ \\
\hline
\multicolumn{4}{c}{$Z=0.008$ $Y=0.25$} \\
\hline
22&10.90 & 4.23&4.35\\
24&10.34  & 4.15 & 4.34\\
26.5&9.73 & 4.07&4.33\\
28& 9.40&4.02& 4.32\\
30&9.04& 3.97& 4.31\\
35& 8.30 & 3.84& 4.29\\
40&7.74& 3.74& 4.28\\
45&7.26& 3.64& 4.26\\
50&6.88& 3.56& 4.25\\
60&6.28& 3.42& 4.23\\
70&5.84& 3.31& 4.21\\
80&5.49& 3.21&4.20\\
90&5.20& 3.13& 4.18\\
100&4.95&3.05 &4.17\\
110&4.75& 2.99 & 4.16\\
120&4.56& 2.92 & 4.15\\
130&4.41& 2.87&4.14\\
140&4.27& 2.81 & 4.13\\
150&4.15& 2.77&4.12\\
160&4.02& 2.71& 4.11\\
170&3.93&2.68& 4.10\\
180& 3.84& 2.64&4.10\\
190&3.75 &2.60&4.09\\
200& 3.66& 2.56& 4.09\\
210&3.60& 2.53&4.08\\
220&3.52 & 2.50&.4.07\\
230& 3.46 &2.46&4.07\\
240&3.39 & 2.43& 4.06\\
250& 3.34 &2.41&4.06\\
\hline
\multicolumn{4}{c}{$Z=0.008$ $Y=0.35$} \\
\hline
20&8.97&4.11&4.36\\
22&8.46& 4.02&4.35\\
24&8.03 &3.95&4.34\\
28&7.38& 3.83&4.32\\
30&7.10&3.77&4.32\\
35&6.53& 3.65&4.30\\
40&6.09&3.54&4.28\\
45&5.75& 3.46& 4.26\\
50&5.46& 3.38&4.25\\
60&4.99& 3.24&4.23\\
70&4.65&3.13&4.21\\
80&4.37&3.03&4.20\\
90&4.15&2.95&4.18\\
100&3.96&2.88&4.17\\
110&3.80&2.81&4.16\\
120&3.66&2.75&4.15\\
130&3.54&2.69&4.14\\
140&3.43& 2.65&4.13\\
150&3.34&2.60&4.12\\
\hline
\hline
\end{tabular}
\end{table}

\subsection{Stellar population models}
\label{sec:popmodels}

To simulate the properties of stellar populations and of synthetic
Cepheids, we used the stellar population synthesis code SPoT,
described in
\citet{Brocato+99a,Brocato+00,Raimondo05,Raimondo09}. Here, we recall
that the code starts directly from stellar evolutionary tracks and
relies on the Monte Carlo techniques for populating the initial mass
function (IMF).  Since our purpose is to have a reasonable number of
variables in the instability strip, we simulated only stars with mass
greater than 1\,M$_\odot$ and assumed a Salpeter IMF \citep{salpeter}
with exponent $\alpha =2.35$. The cut at low mass stars does not
affect the conclusions of this paper because we are interested to
bright high mass stars evolving off the main sequence.  All the
evolutionary phases, from the main sequence (MS) up to the asymptotic
giant branch, are covered by models. For each set of simple stellar
population (SSP) parameters (i.e., age, $t$; metallicity, $Z$; and He
abundance, $Y$), 100 independent simulations each containing 5000
stars (in the mass range 1-12\,M$_\odot$) are computed. Each single
simulation corresponds to a stellar population having a total mass of
the order of $\sim10^4$\,M$_\odot$, including low mass stars down to
0.1\,M$_\odot$.

We computed synthetic magnitudes and colours (in the $U,B,V,R,I,J,H$
and $K$ bandpasses) by adopting \cite{castelli} stellar atmospheres
library. The code has been implemented by including specific routines
to compute the number of predicted Cepheids and, for each variable,
the pulsation mode and period \citep{carini}. For populations with
$Y=0.25$, we adopted the instability strip calculated by \citet{bms99}
and \citet{b01}, for $Y=0.35$ we used the values published in
\citet{carini}. We consider ages from 20\,Myr up to about 250\,Myr
(with an age step of $\sim2$\,Myr up to 50\,Myr, and $\sim10$\,Myr
up to 250\,Myr) for the first stellar generation ($Y=0.25$), and from
20\,Myr up to 150\,Myr (with the same age steps)
for Cepheids belonging to the second generation of stars ($Y=0.35$).
For older ages no Cepheids are found in the populations (see Fig.~\ref{fig1}).
The age steps are chosen to reproduce in detail the main features of
the Cepheid populations as a function of the age. The mean mass,
luminosity and effective temperature of the stars at the termination
of the Main Sequence luminosity function, for each population, are
listed in Table~\ref{to}.

In Fig.~\ref{fig1}, we present the brightest part of the
Hertzsprung-Russel (HR) diagrams expected from a sample of stellar
populations having He abundances $Y=0.25$ (left panel) and $Y=0.35$
(right panel) and selected ages.
The number of stars in each synthetic model shown in Fig.~\ref{fig1}
is quite high to ensure that the He-burning phase is well populated
also when computing very young stellar populations. This is obtained
by collecting together all the 100 independent simulations, in this
way each plotted model represents a single burst stellar population
with a mass of $\sim10^6$\,M$_\odot$. The instability strips
derived from the pulsation models are also plotted: the blue short-dashed line is
the first overtone blue edge (FOBE), the blue solid line is the
fundamental blue edge (FBE), the red dotted line is the first overtone
red edge (FORE) and the red solid line is the fundamental red edge
(FRE). Green circles are the stars predicted to be variables.

Within the theoretical framework described in the previous sections,
we are able to put constraints on the oldest stellar populations where
Cepheids can be observed. In the case of $Y=0.25$ (left panel) the
oldest synthetic population expected to host Cepheids is found to be
254\,Myr old, while for $Y=0.35$ (right panel) the Cepheids should not
be observed for ages older than 156\,Myr. The exact age values are
obtained by computing synthetic models with a very narrow step in age.
The difference in the age limits is due to the fact that stars with a
high He abundance, due to the higher molecular weight, evolve in a
shorter time and are more luminous and hotter than stars with a lower
He abundance at fixed mass. Therefore, the population with $Y=0.35$
and age $t\sim150$\,Myr contains stars with mass lower than
$\sim3$\,M$_{\odot}$ as well as a population with age $\sim250$\,Myr
but with $Y=0.25$ (Table~\ref{valori}). The reason for the age limits
is well known. In old population, the hotter extremity of the blue
loop of the evolving stars becomes cold enough to lie outside of the
instability strip, so that no star falls inside the instability strip
and no Cepheid is foreseen, as clearly shown in Fig.~\ref{fig1}.  This
implies that there are no Cepheids with $Y=0.35$ pulsating in the
fundamental mode with $\log P \lesssim 0.50$ and no Cepheids with
$Y=0.25$ at $\log P \lesssim 0.2$. In Table~\ref{valori} we report the
maximum age of the stellar system hosting Cepheids, and the physical
properties of pulsating stars, namely mass, luminosity, temperature
and period for the two analysed He contents.

\begin{table}
\centering
\caption{
  Mean values for key quantities of variables in the oldest stellar
  population where Cepheids are expected, for the two He contents.}
\label{valori}
\begin{tabular}{cccccc}
\hline  
$Y$ & $t$ (Myr) & $M$/M$_\odot$ & $\log L/L_{\odot}$ & $\log T_{eff}$ & $\log P$ \\
\hline
0.25  & 254 & 3.60 & 2.67 & 3.76 & 0.28 \\
0.35  & 156 & 3.56 & 2.97 & 3.75 & 0.57 \\
\hline
\end{tabular}
\end{table}

\begin{figure*}
\center
\includegraphics[trim= 0.8cm 0.8cm 0.8cm 0.8cm, clip=true,width=.9\columnwidth]{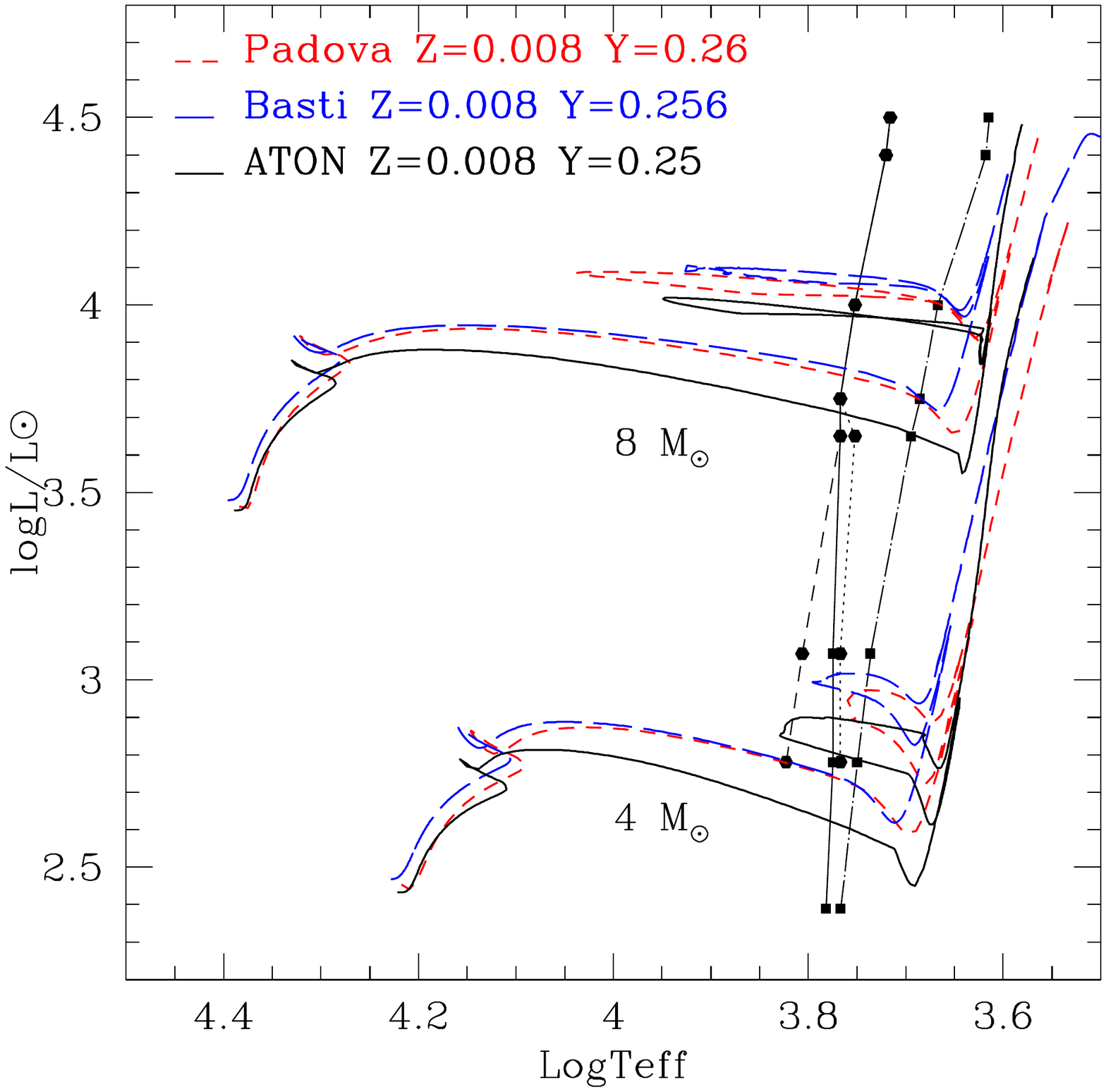}
\includegraphics[trim= 0.8cm 0.8cm 0.8cm 0.8cm, clip=true,width=.9\columnwidth]{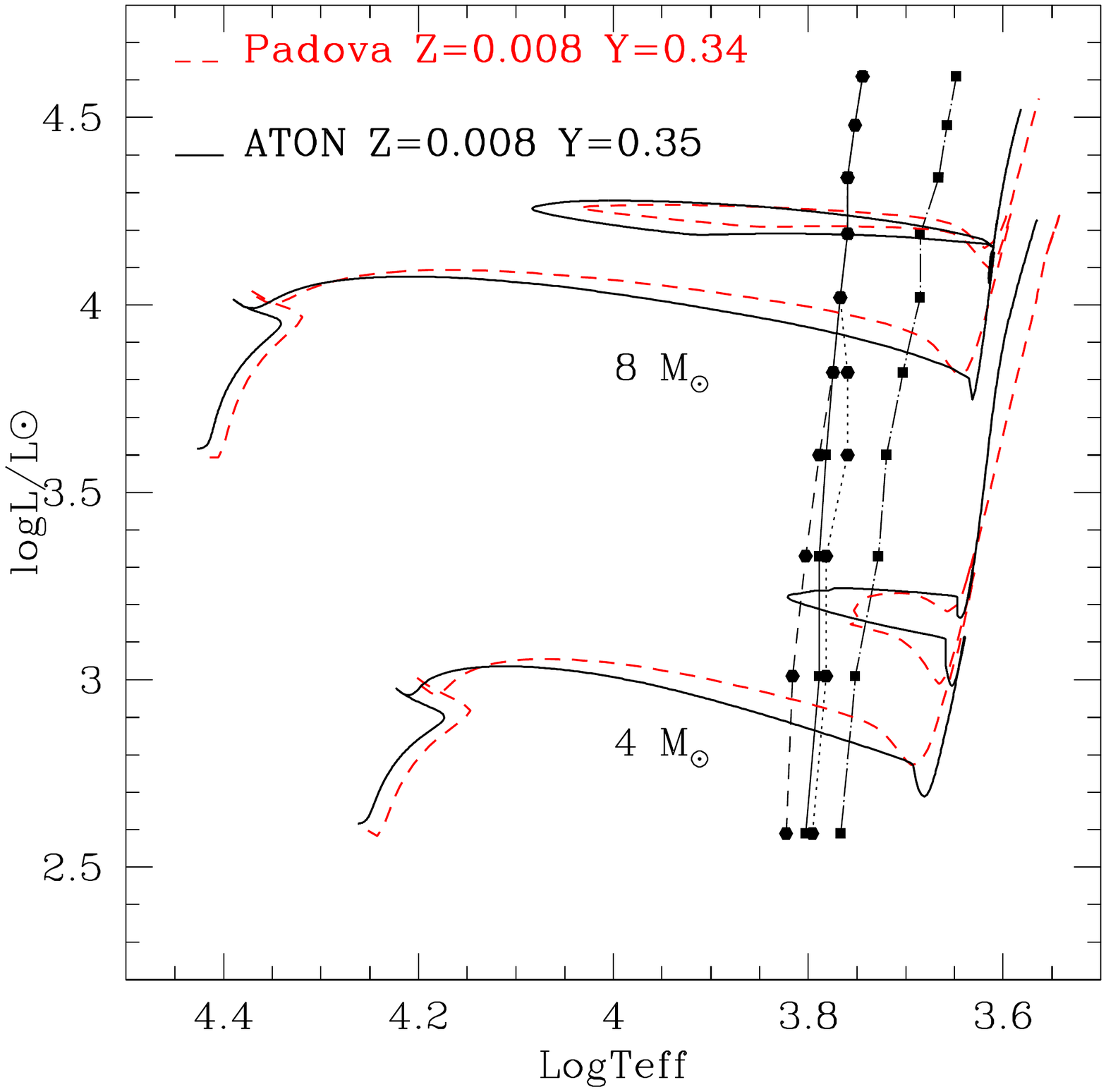}
\vspace{-1cm}
\caption{The evolutionary path in the HR diagram of a 4\,M$_{\odot}$ and a 8\,M$_{\odot}$
  model calculated with the ATON code (black solid line) is compared to that from other
  databases: Basti (long dashed blue line) and Padova (short dashed
  red lines) for $Y=0.25$ (left panel) and $Y=0.35$ (right panel, see text). The
  FRE (dot dashed black line), FORE (dotted black line), FBE (solid
  black line) and FOBE (short dashed black line) derived
  in \citet{carini} are also shown.
 }
\label{tracce}
\end{figure*}

We point out that our results depend on the evolutionary scenario we
have adopted. Different assumptions concerning the surface boundary
condition, solar calibrated mixing length, model of convection,
chemical composition, equation of state (EOS), gravitational settling,
mass loss and other physics, may result in differences in the
evolutionary path in the HR diagram. To throw some light on this
point, in Fig.~\ref{tracce} we compare our tracks (solid black lines)
with the ones from other popular databases: the Padova
database\footnote{\url{http://pleiadi.pd.astro.it/}} (short dashed red
lines, $Z=0.008, Y=0.26,0.34$), and the Basti
database\footnote{\url{http://basti.oa-teramo.inaf.it/}} \citep[long
dashed blue lines, $Z=0.008, Y=0.256$; see][for
details]{bertelli09,pietri04,pietri06}. We outlight that the $Z$ and
$Y$ values are not exactly the same, but are chosen as similar as
possible, when available. In Fig.~\ref{tracce} we plot the
evolutionary tracks for models of 4\,M$_\odot$ and 8\,M$_\odot$, for
the labelled He abundance values. The black vertical lines represent
the instability strip boundaries.

First, we can see that the effect of He abundance is fairly similar in
our and Padova tracks. The ATON (black) He-rich 4\,M$_\odot$ model is
more luminous of $\sim0.2$\,dex at the TO and of $\sim0.3$\,dex at the
He-burning loop than the standard one; for the 8\,M$_\odot$ model the
overluminosity values are 0.15\,dex and 0.25\,dex, respectively. The
extension of the loop for the 4\,M$_\odot$ models is similar for the
two He abundances, while for the 8\,M$_{\odot}$ models the He-enhanced
path is more extended in temperature of about 0.15\,dex. The Padova
tracks show a similar behaviour in luminosity while the extension of
the loop remains comparable for the two He values. In the Basti
database there are not He-enhanced tracks for intermediate-mass stars
in the He burning evolutionary phase.

For what concerns differences in the stellar evolution codes and in
the adopted input physics we focus on He-standard models (left panel),
because they are available in each database (even with slightly
different $Y$ values), making possible a complete comparison. The left
panel of Fig.~\ref{tracce} shows that
the tracks slightly differ in luminosity ($\Delta \log L/L_\odot$
$\thickapprox 0.1$\,dex) and in
the extension of the core-He burning loop ($\Delta \log T_{eff}
\lesssim 0.1$\,dex). For example, the 4\,M$_{\odot}$ ATON model (black
line) is less luminous but it has a blue loop more extended than the
others.  The behaviour in luminosity is the same for the model with
8\,M$_{\odot}$, but in this case the Padova track extends to hotter
temperatures.
A deep analysis of the impact of different evolutionary codes and
input physics on the evolutionary behaviour of stellar models is
beyond the aim of the present work, and we refer the interested reader
to \citet[e.g.][]{chiosi1982,castellani1990,ventura98,ventura2005}.
The comparison illustrated above suggests that the number and the
thermodynamic properties of the synthetic Cepheids may slightly change
according to the used set of stellar models.  However, the differences
are expected to affect only marginally the main results obtained in
this work.  Consequently, the synthetic PLRs and PWRs based on
evolutionary tracks by other authors show slopes and zero points
similar to ours, within the uncertainties. This is the case of results
by \cite{b10}, whose relations are in very good agreement with ours,
even taking into account that their Wesenheit relations are calculated
by adopting $R_V=3.23$ instead of $R_V=3.1$.  The PLRs and PWRs by
\cite{caputo00} and \cite{fio2007} are also in agreement with our
results, when statistical effects in the number of Cepheids are taken
into account (see below Tab. \ref{tabl1} in
Sec.~\ref{sec:popeffects}).

\section{He-enhanced Cepheids and comparison with empirical relations.}
\label{sec:observation}

Our models make possible to simulate the expected number and
properties of Cepheids in stellar systems.  In particular, we are
interested to investigate the observational features of Cepheids in
stellar populations having a metallicity similar to the typical value
of the LMC ($Z\sim0.008$) and different He abundances: $Y=0.25$, which
represents the classical case, the higher value $Y=0.35$, and a simple
mixture of the two stellar populations.  As a first step, we assembled
together all the simulations computed assuming $Z=0.008$ and $Y=0.25$,
considering all the ages. The resulting large sample of Cepheids
corresponds to that of a stellar population generated from a series of
close star-formation bursts with short age steps (2-10\,Myr). It is
worth noting that, this can be considered as a stellar population
whose stars are generated according to a constant Star Formation Rate
(SFR) from 20\,Myr to 250\,Myr.  We tested this assumption by
computing a stellar population model in which a continue star
formation is actually assumed in the same age interval. We found that
the PLRs and the PWRs are in full agreement with those computed from
the simulation described above, where the constant star formation has
been reproduced by the series of SSPs with the adopted age steps. This
is a well known result in stellar population synthesis, as a complex
stellar population can always be expanded in a series of simple
stellar populations \citep{renzini86}.

The same assembling procedure has been performed for $Y=0.35$ models
with age ranging from 20 to 150\,Myr.  We ended up with two stellar
systems having a total mass in stars of $\sim10^8$M$_\odot$, and a
total number of predicted Cepheids of $\sim$\,10000 for $Y=0.25$ and
$\sim$\,7700 for $Y=0.35$.

From the analysis of the two samples, we derived the PLRs and the PWRs
in different photometric bands, namely in $B$, $V$, $I$, $J$, and $K$,
and $W(B,V)$, $W(V,I)$, $W(V,K)$, $W(J,K)$. The mean slopes and
intercepts are reported in Table~\ref{tablemean}.

These theoretical relationships will be compared with the empirical
ones, derived by observed samples of LMC Cepheids, in the following
subsections.

\begin{table}
\centering
\caption{
  Theoretical PLRs and PWRs  for fundamental classical Cepheids
  derived from a linear fit: $M_{\lambda}$ (or $W_{bands})= \beta \times \log P + \alpha$.
  The standard deviation of the slopes $\sigma_{\beta}$ and the intercepts 
  $\sigma_{\alpha}$
  are also reported. }
\label{tablemean}
\begin{tabular}{ccccc}
  \hline
  Band & $\beta$ & $\sigma_{\beta}$ & $\alpha$ & $\sigma_{\alpha}$      \\
  \hline
  \multicolumn{5}{c}{$Z=0.008$ $Y=0.25$} \\
  \hline
  B & -2.598  & 0.006 & -0.638  & 0.003\\
  V & -2.836  & 0.005 & -1.132  & 0.002 \\
  I &  -3.003 & 0.003 & -1.735  & 0.001 \\
  J & -3.137  & 0.003 & -2.138  & 0.001\\
  K &  -3.252 & 0.002 & -2.471 & 0.001 \\
  W(B,V) & -3.574 &0.002 & -2.664  &  0.001 \\
  W(V,I) & -3.262 & 0.002&  -2.663 & 0.001 \\
  W(V,K) & -3.306 & 0.001& -2.645  & 0.001 \\
  W(J,K) & -3.318 & 0.001& -2.663 & 0.001 \\
  \hline
  \multicolumn{5}{c}{$Z=0.008$ $Y=0.35$} \\
  \hline
  B & -2.591 & 0.015 & -0.798 &  0.011 \\
  V & -2.859 &0.011 &   -1.172 & 0.008 \\
  I & -3.053 &  0.008& -1.680  & 0.006\\
  J & -3.195 &  0.006 & -2.022 &  0.004 \\
  K & -3.319 & 0.004 & -2.295    &0.003 \\
  W(B,V) & -3.691 &0.003 & -2.332 & 0.002 \\
  W(V,I) &-3.350 &0.003 & -2.463 &  0.002  \\
  W(V,K) &-3.378& 0.003 & -2.441 & 0.002\\
  W(J,K) &-3.390&0.003& -2.452 & 0.002 \\
  \hline
\end{tabular}
\end{table}

\subsection{PL relations}

Our PLRs in the $V$, $I$, $J$ and $K$ bands are compared to those
derived by \cite{storm}, who analysed 22 Cepheids pulsating in the
fundamental mode in the LMC.  The authors selected stars which
simultaneously have high-quality near-infrared light-curves from
\cite{persson04} and very accurate optical photometry from OGLE-III
\citep{udalski08,sosz08}.  Storm et al. determined the following
relations from a linear regression to the absolute magnitudes and
$\log P$, from the infrared surface brightness analysis:

\begin{equation}
M_V = -2.78 \, (\log P-1)-4.00
\end{equation}
\vspace{-.2cm}
\begin{equation}
M_I=-3.02 \, (\log P-1)-4.74
\end{equation}
\vspace{-.2cm}
\begin{equation}
M_J=-3.22 \, (\log P-1)-5.17
\end{equation}
\vspace{-.2cm}
\begin{equation}
M_K=-3.28 \, (\log P-1)-5.64
\end{equation}

The dispersion around the fits is 0.26\,mag in the $V$ band, and
0.21\,mag in the others.  Note that their relations are in good
agreement with those of \cite{persson04} and \cite{ripepi12} in the
near-infrared bands, and with \cite{udalski00} and \cite{sosz08} in
the optical ones.

\begin{figure}
\center
\includegraphics[trim= .1cm 1cm 1cm 2cm, clip=true,width=.9\columnwidth]{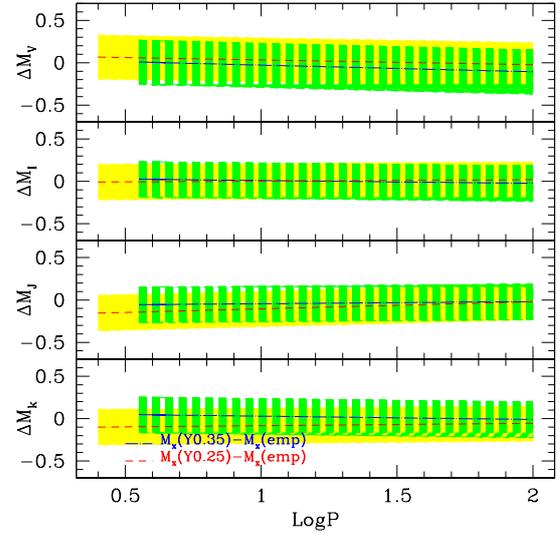}
\vspace{-1cm}
\caption{
Differences between our theoretical and empirical PLRs in the $K$, $J$, $I$ and $V$ bands
by \protect\cite{storm}.
The 1\,$\sigma$ uncertainties are reported as filled yellow ($Y=0.25$) and shaded green ($Y=0.35$) areas. }
\label{pl2}
\end{figure}

The relations above are in very good agreement within the
uncertainties with our determinations (Table~\ref{tablemean}), as
shown in Fig.~\ref{pl2}, where the difference between the empirical
and theoretical PLRs is reported as function of the photometric band.

The first thing that stands out is that the relations derived from
models with the two considered \emph{pure} He abundances are very
similar. In each band both the theoretical relations are in agreement
with the empirical ones within the uncertainty, but a better agreement
is found when assuming the standard He abundance.  In the $K$ band the
relation obtained from models with $Y=0.35$ diverges from the observed
ones at $\log P < 1.5$, with a maximum difference in magnitude of
$\sim0.05$\,mag at $\log P=0.6$. The opposite trend is found in the
$V$ band, where the difference can be as high as 0.1\,mag at $\log
P>1$, in agreement with the results of \cite{fio02}. The $I$ and
$J$-band differences are negligible ($\leq0.1$\,mag).  In conclusion,
the uncertainties due to the Helium abundance are of the order of 7\%
in $B$, 4\% in $V$, 1\% in $I$, 1\% in $J$ and 4\% in $K$.

\begin{figure}
\center
\includegraphics[trim= .1cm 1cm 1cm 2cm, clip=true,width=.9\columnwidth]{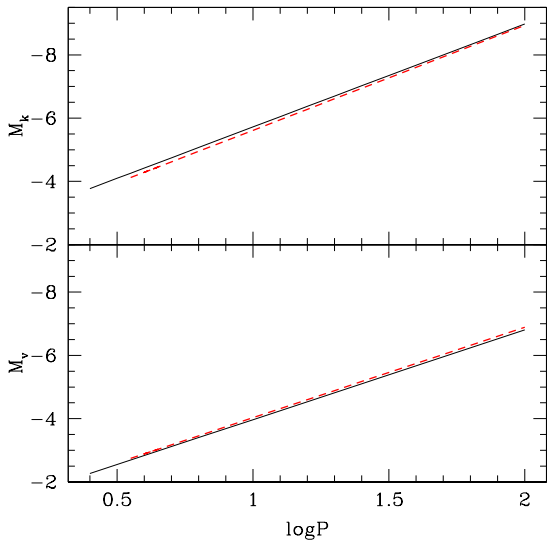}
\vspace{-1cm}
\caption{Comparison between the $V$ (bottom panel) and $K$ (top panel) relations with
  $Y=0.25$ (black solid line), $Y=0.35$ (red dashed line).
  }
\label{fitkv}
\end{figure}

The second thing that stands out is that the blue long dashed line,
representing the difference between the theoretical relation with
$Y=0.35$ and the empirical one, moves from negative to positive values
passing from the $V$ to $K$ band.  This behaviour is more evident in
Fig.~\ref{fitkv} where the PLRs in $V$ and $K$ bands for both He
abundances are compared.

The $\log P$-$V$ relation with $Y=0.35$ stands at brighter $V$
magnitude for all periods, and the two lines are almost parallel each
other. The $\log P$-$K$ relations have the opposite behaviour and
their magnitude difference decreases with the period.  This happens
because variable stars with the same period but different He abundance
have very similar luminosity and surface gravity but different mass
and effective temperature. In general, at fixed age, the He-rich stars
are less massive (because their evolution is more rapid) but hotter
than stars with standard He abundance.  Since the bolometric
correction in $V$ ($BC_V$) decreases with effective temperature, while
$BC_K$ increases, it turns out that the behaviour in the $\log P$-mag
relations is opposite.  To clarify this point, we show in
Table~\ref{tab6} the thermodynamic values, the magnitudes in the $V$,
$I$ and $K$ bands and the bolometric corrections of two couple of
variables with same period and age.  We have chosen two stars at
130\,Myr and 30\,Myr.

\begin{table*}
\centering
\small
\caption{
  Thermodynamic values, magnitudes in the $V$, $I$ and $K$ bands
  and the bolometric corrections of two couple of stars at $\log P$=0.6 and 1.59,
  and age $t=130$\,Myr and 30\,Myr, respectively.}
\label{tab6}
\begin{tabular}{cccccccccccccc}
  \hline
  $Y$&$t$&M/M$_\odot$&$T_{eff}$ & $\log( L/ L_\odot)$ & $\log g$ & $\log P$ & $V-K$ & $ V$ & $ I$ & $K $ & $BC_V$ & $BC_I$ & $BC_K$ \\
  & (Myr) & & (K) & & & &(mag) &(mag)& (mag)&(mag)& (mag)&(mag)& (mag)\\
  \hline
  0.25	&130&4.67&5495& 3.06&1.97&0.60& 1.65& -2.77&-3.50 &-4.42& -0.13&0.60 & 1.51\\
  0.35	&130&3.84&5888& 3.10 &1.95 &0.60&1.37& -2.94& -3.55&-4.31 &-0.06&0.55&1.31\\
  0.25	&30&9.78&4897&4.19&0.93&1.59&2.22&-5.41&-6.37 &-7.63&-0.32&0.64 &1.90\\
  0.35  &30&7.82&5128&4.19&0.91&1.59&2.00&-5.50&-6.38 &-7.50&-0.23& 0.65&1.77\\
  \hline
\end{tabular}
\end{table*}

We can see the two stars in each couple mainly differ in mass and
effective temperature.  Higher is the temperature, higher is $BC_K$,
but lower is $BC_V$; $BC_I$ shows an intermediate behaviour.  In the
first case ($t=130$\,Myr) the temperature difference between the two
stars is about 390\,K, this implies a $\Delta BC_V = -0.07$\,mag,
$\Delta BC_I= 0.05$\,mag, and $\Delta BC_K= 0.2$\,mag.  In the second
case ($t=30$\,Myr), with a $\Delta T_{eff}\sim230$\,K, the differences
in the bolometric corrections are $\Delta BC_{V}=-0.09$\,mag, $\Delta
BC_I=0.01$\,mag and $\Delta BC_K=0.12$\,mag.  Therefore, by increasing
$\Delta T_{eff}$, $\Delta BC_K$ increases, but $\Delta BC_V$
decreases.  For this reason, the two relations $\log P$-$K$ are
inverted with respect to the $\log P$-$V$ relations, and their
magnitude difference is higher at shorter periods than at longer ones.\\

In conclusion, we confirm that the He content has an effect not
detectable on the PLRs with the sensitivity of the current
instruments, so that we cannot discriminate whether the population is
He-enhanced or not.  The $\log P$-$V$ and $\log P$-$K$ relations show an
opposite behaviour which depends on the difference in the effective
temperature between the variables having same period but different He.
This effect will be accentuate in the Wesenheit relations, which
depends on the colour of the stars (see next section).

\subsection{Wesenheit relations}
\label{wasenrel}

We compare our theoretical Wesenheit relations with the ones derived
by \cite{ripepi12} and \cite{inno}. The former authors analysed
$K_s$-band light curves of the Cepheids in the LMC observed by the
VISTA Magellanic Cloud survey\footnote{Based on observations made with
  ESO Telescopes at the La Silla or Paranal Observatories under
  programme ID(s) 179.B-2003(D), 179.B-2003(C), 179.B-2003(B)}
\citep[VMC,][]{cioni11,ripepi12}. The stars are detected in the $Y$,
$J$ and $K_s$ filters; the Cepheid $ K_s$ light curves are very well
sampled, with at least 12 epochs, and with typical errors of 0.01
mag. They mapped two fields centred on the south ecliptic pole and the
30 Doradus star-forming regions, respectively. We use only their
results from the second field, because the sample is the richest one,
being composed by 172 classical Cepheids pulsating in the fundamental
mode and 150 in the first-overtone. Since the longest period in the
data-set is 23 days, Ripepi and collaborators complemented the sample
with literature data, including other 80 Cepheids pulsating in the
fundamental mode from \cite{persson04}. From all data they derived:

\begin{displaymath}
W(V,K)= -3.325 \, \log P + 15.870
\end{displaymath}
with a dispersion of 0.078. No correction for the inclination of the
LMC disc was applied.

\begin{figure}
\center
\includegraphics[trim= .1cm 1cm 1cm 2cm, clip=true,width=1.0\columnwidth]{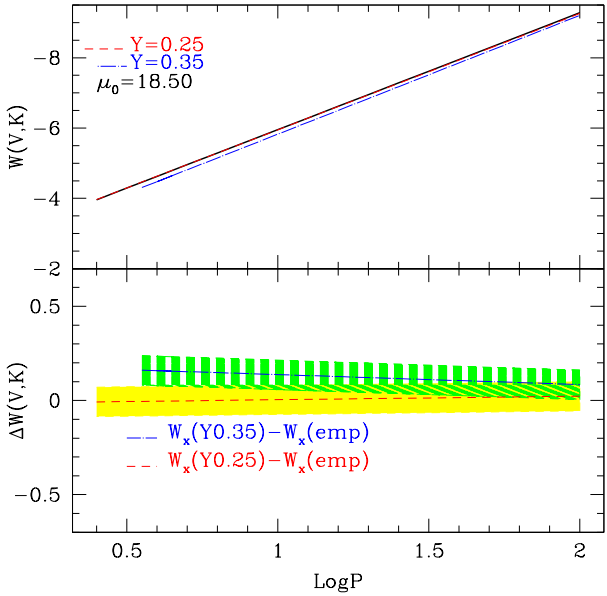}
\vspace{-2cm}
\caption{Upper panel: Comparison between the $W(V,K)$ relation (black
  solid line) of \protect\cite{ripepi12} and the theoretical ones: the
  red short-dashed line corresponds to $Y=0.25$ and the blue dot-long
  dashed line to $Y=0.35$ (see Table~\ref{tablemean}). Lower panel:
  Difference between theoretical and empirical Wesenheit relations
  plotted in the upper panel.  The used distance modulus of the LMC is
  18.50\,mag \citep{ripepi12}.  Symbols and colours are as in the
  previous figure.
 }
\label{wvk}
\end{figure}

We plot the comparison of the empirical relation with the present
results in the top panel of Fig.~\ref{wvk}. The theoretical relation
obtained with the standard He abundance (red short-dashed line)
overlaps the observed relation (black solid line) over the full range
of periods, as clearly seen in the lower panel of the figure, where
the difference between the empirical and theoretical PWRs is
shown. The red line is nearly constant because the two relations are
similar (throughout the paper, we assume the LMC distance modulus of
$\mu$=18.50), the small difference between the slopes of the two
relations (0.019) leads to a maximum difference $\Delta W(V,K)$ of
about 0.02\,mag for $\log P >1.8$. Instead, the PWR derived from
He-enhanced models (blue dot-long-dashed line) is tilted with respect
to the observational one: the difference between the slopes is large
reaching nearly 0.15 mag at short periods ($\log P\sim0.6$), because
the He-enhanced models predict fainter magnitudes for the periods
considered in this work. As we have already mentioned,
the shortest period reached is $\log P\sim0.5$, because in stellar
populations older than about 150 Myr stars do not cross the
instability strip during their evolution, therefore there are not
Cepheids with $Z=0.008$, $Y=0.35$ and $\log P < 0.5$.

\begin{figure}
\center
\includegraphics[trim= .1cm 1cm 1cm 2cm, clip=true,width=1.0\columnwidth]{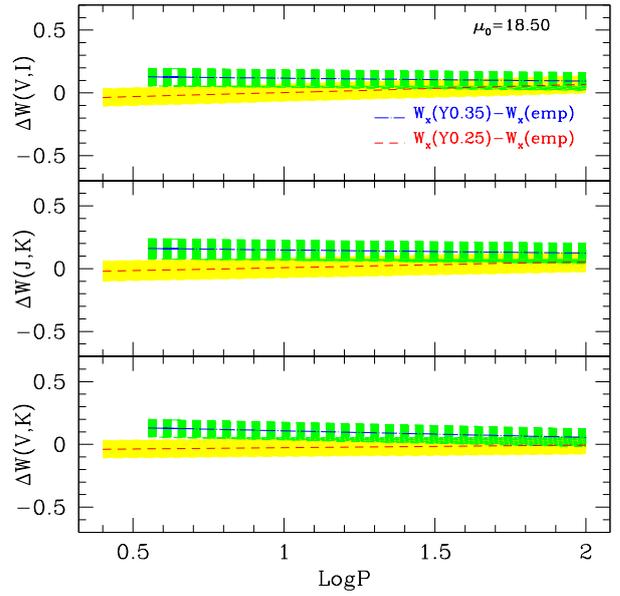}
\vspace{-2cm}
\caption{Difference between theoretical and empirical PWRs
  by \protect\cite{inno} in the $(V,I)$, $W(J,K)$ and $(V,K)$ bands. Symbols
  and colours are as in the previous figure.
  }
\label{deltawas}
\end{figure}

To investigate this issue in other photometric bands, we consider the
PWRs published by \cite{inno}. They combined the data in the $V$ and
$I$ bands of the OGLE III (Optical Gravitational Experiment)
catalogue\footnote{\url{http://ogledb.astrouw.edu.pl/~ogle/CVS/}} \citep{sosz08} with
data in the $J$, $H$, and $K$ bands from the near-infrared catalogue
of IRSF/SIRIUS Near-Infrared Magellanic Clouds Survey provided by
\citet{kato}, ending up with a sample of 1840 fundamental
classical-Cepheids in the LMC. From linear regressions the authors
found:

\begin{equation}
W(V, K)= -3.326 \, \log P + 15.901
\end{equation}
\vspace{-.2cm}
\begin{equation}
W(V, I ) = -3.327 \, \log P + 15.899
\end{equation}
\vspace{-.2cm}
\begin{equation}
W(J,K)=-3.365 \, \log P +15.876
\end{equation}
The dispersion obtained is 0.07\,mag for $W(V,K)$ and $W(V,I)$, and
0.08\,mag for $W(J,K)$.

Again, the good agreement between our theoretical standard
He-abundance and the observational Wesenheit relations is evident in
Fig.~\ref{deltawas}. Small differences arise in the zero points and
the slopes, up to $\sim-0.03$\,mag for $W(V,K)$ (lower panel), while
from nearly $-0.02$ to 0.06\,mag for $W(J,K)$ (middle panel) and from
nearly $-0.04$ to 0.08\,mag for $W(V,I)$ (top panel), i.e. our
standard relations are in agreement with the empirical ones within the
errors over the full range of periods.  In the case of $Y=0.35$ the
relations are tilted with respect the observational ones, the
difference reaching values larger than $\sim0.15$\,mag. This result
implies that, if all stars in a generic stellar population have an
enhanced He abundance, the uncertainty in the evaluation of the
distance will be as high as 0.1-0.15\,mag. To enlighten this point,
let us suppose that one is observing a sample of Cepheids all having a
He abundance $Y\sim0.35$. If the He-standard Wesenheit relations are
used, it leads to a systematic error of the order of 3-7\% in
evaluating the distance modulus, with the error increasing when $\log
P$ decreases. For example, in the case of W(V,I), the systematic error
is about 3\%.

\subsection{Mixed Population}
\label{sec:mixed}

\begin{table}
\centering
\caption{
Theoretical PLRs and PWRs for fundamental classical Cepheids derived from a linear fit for a
mixed population containing stars with $Y=0.25$ and $Y=0.35$ ($Y_{MIXED}$):
$M_{\lambda}$ [or $W({band_1,band_2)}]= \beta\times\log P + \alpha$. The standard deviation of the slopes
$\sigma_{\beta}$ and the intercepts $\sigma_{\alpha}$ are also reported.}
\label{tablemix}
\begin{tabular}{ccccc}
\hline
Band & $\beta$ & $\sigma_{\beta}$ & $\alpha$ & $\sigma_{\alpha}$      \\
\hline
\multicolumn{5}{c}{$Z=0.008$ $Y_{MIXED}$} \\
\hline
 B & -2.784  & 0.005 & -0.612  & 0.003\\
 V & -2.909  & 0.004 & -1.120  & 0.002 \\
 I & -2.989 & 0.003 & -1.733  & 0.002 \\
 J & -3.057  & 0.002 & -2.146  & 0.001\\
 K &  -3.109 & 0.002 & -2.487 & 0.001 \\
 W(B,V) & -3.296 &0.003 & -2.695  &  0.002 \\
 W(V,I) & -3.113 & 0.002&  -2.678 & 0.001 \\
 W(V,K) & -3.135 & 0.002& -2.664  & 0.001 \\
 W(J,K) & -3.139 & 0.002& -2.683 & 0.001 \\
\hline
\end{tabular}
\end{table}

Before concluding the section, we perform a numerical exercise to
study the case where the two populations are mixed together. Since our
models foresee 9918 Cepheids for $Y=0.25$ and 7641 for $Y=0.35$, the
contamination of He-enhanced stars is of the order of 40\%.
We test this population split, because it can be considered one of the
worst case scenarios, useful to derive an upper limit of the
uncertainty on the galaxy distances caused by different Helium
abundances in the galaxy field populations.  In addition, it may be
interesting for young cluster populations.

\begin{figure*}
 \center
\includegraphics[width=.9\columnwidth]{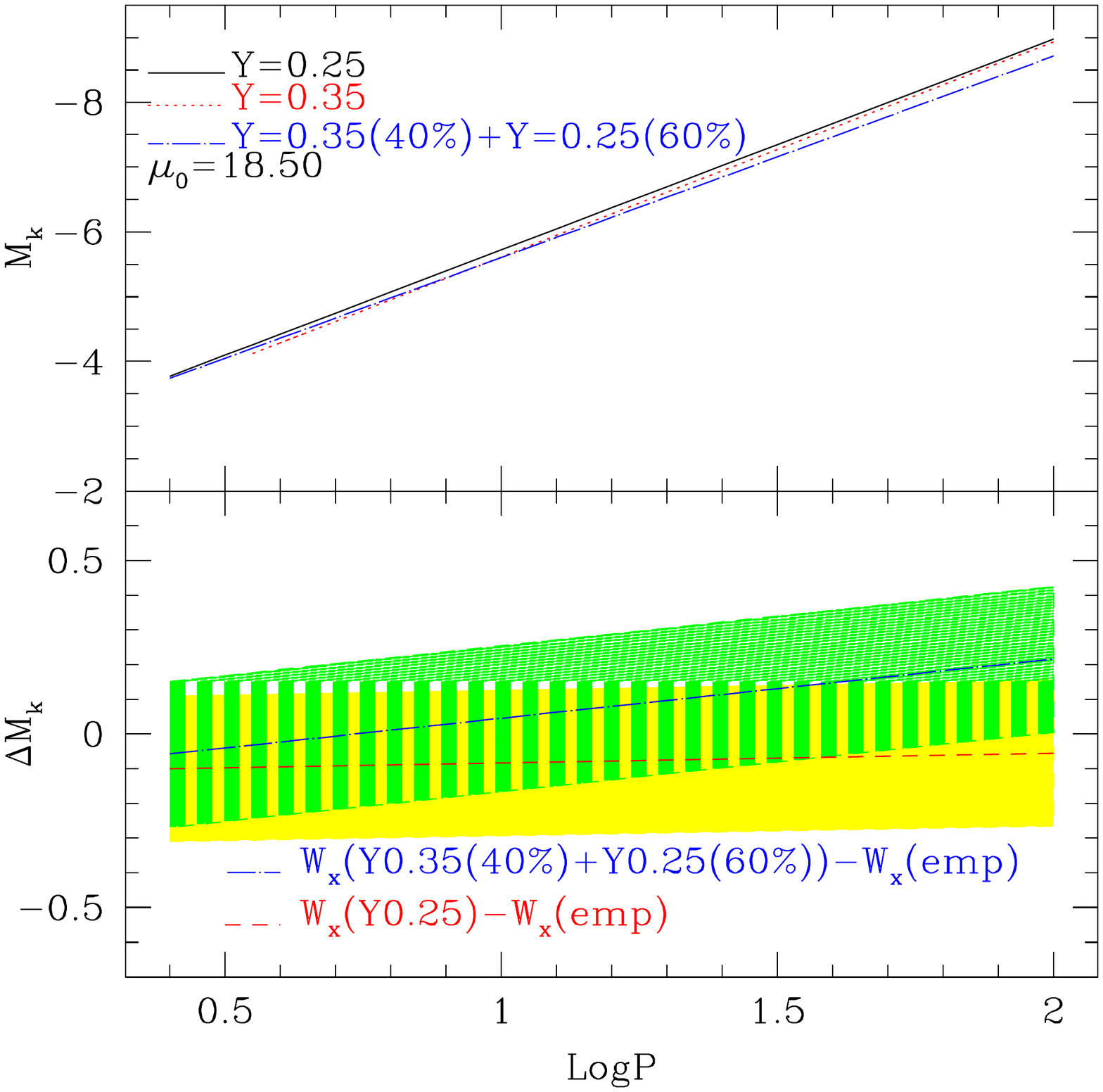}
\includegraphics[width=.9\columnwidth]{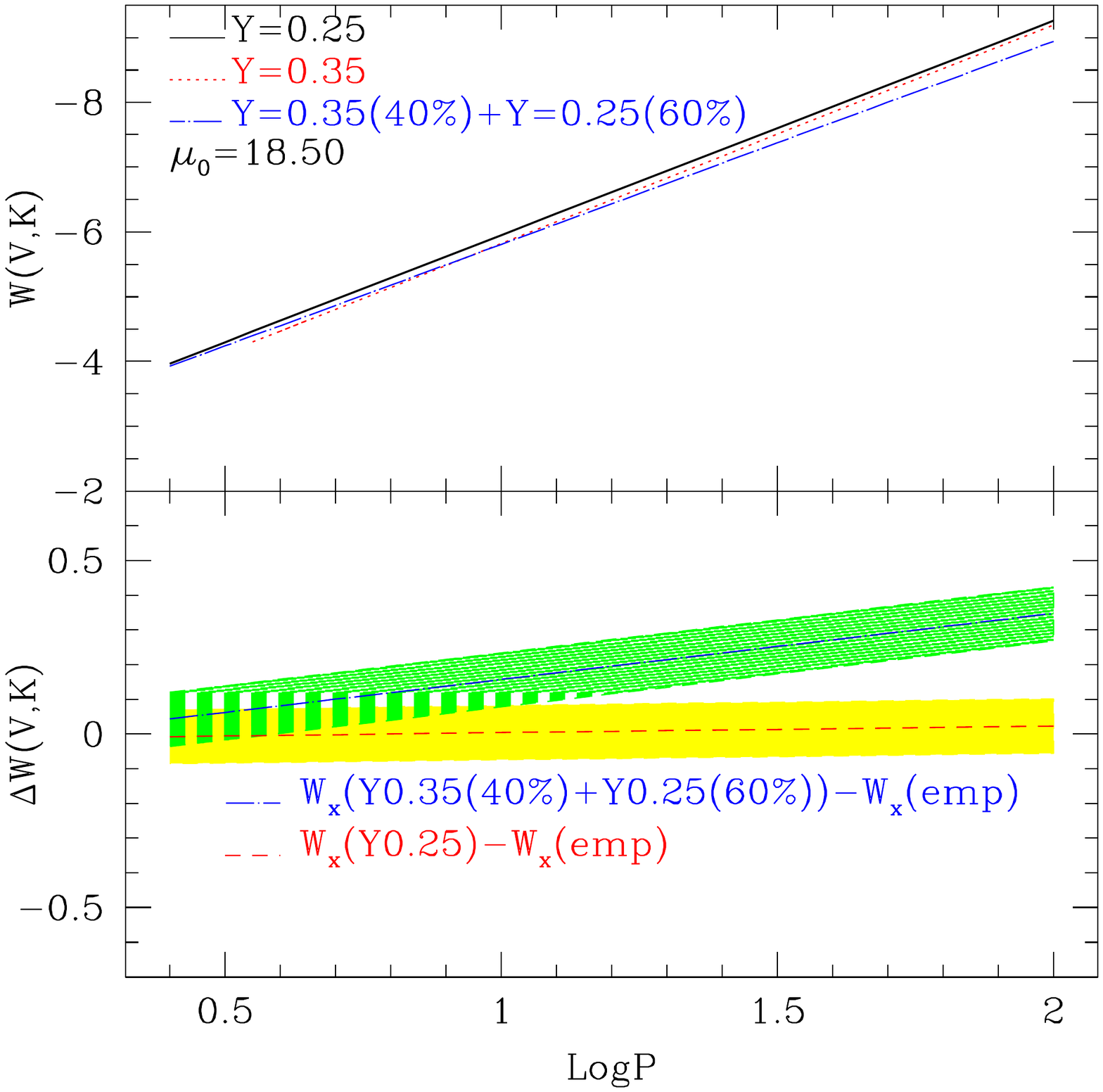}
\vspace{-1cm}
\caption{Upper panels: Comparison between the $P-M_k$ (left panel) and
  $W(V,K)$ (right panel) relations derived from a \emph{pure} $Y=0.25$
  population (black solid line), \emph{pure} $Y=0.35$ population (red
  dotted line), and a mixed population $Y_{MIXED}[60\%(0.25);
  40\%(0.35)]$ (blue dot-long dashed line). Lower panel: The
  differences between theoretical and empirical relations, i.e
  $M_k[Y=0.25]-M_k[empirical]$ (right panel),
  $W(V,K)[Y=0.25]-W(V,K)[empirical]$ (left panel) represented by a red
  short-dashed line, and $M_k[Y_{MIXED}[60\%(0.25);40\%(0.35)]]
  -M_k[empirical]$ (right panel) and
  $W(V,K)[Y_{MIXED}[60\%(0.25);40\%(0.35)]]-W(V,K)[empirical]$ (left
  panel) represented by a blue dot-long-dashed line .  The empirical
  $P-M_k$ relation is from (\protect\citealt{storm} ($P-M_k$), while
  the $W(V,K)$ relation is from \protect\citealt{ripepi12}).  The
  $1$\,$\sigma$ uncertainties are reported as filled yellow ($Y=0.25$)
  and shaded green ($Y_{MIXED}[60\%(0.25);40\%(0.35)]$).}
\label{mix}
\end{figure*}

In this case, the relations are tilted with respect to those with a
\emph{pure} abundance of He, as already shown in \cite{carini}. Here,
we extended the computation and the results on PLRs and PWRs as
reported in Table~\ref{tablemix}. The systematic uncertainty in the
evaluation of distances can be considered negligible only in the $I$
band, because the PLR of the mixed population is similar (within
$\sim1$\%) to the one derived for a \emph{pure} $Y=0.25$
population. The other uncertainties are 12\%, 5\%, 5\%, and 10\% for
the $B$, $V$ $J$ and $K$ bands and decrease when the period decreases.
Since the PLR in the K band is less affected by systematic errors due,
for example, to the finite width of the instability strip, non
linearity and chemical composition \citep[e.g.][]{bccm99,caputo00} or
reddening and intrinsic dispersion \citep[e.g.][]{madore} and to
stochastic effects (see Table \ref{tabl1} in sec. \ref{sec:popeffects}
), the effect of Helium seems to be the potentially the most important
one.  This becomes particularly evident in the case of mixed
population, where the uncertainty due to the Helium effect on the
distance determination can be as high as 10\%. This can be seen in the
upper panel of Fig. \ref{mix}, that shows, the PLRs in the K band for
three different cases: \emph{pure} $Y=0.25$ (black solid line),
\emph{pure} $Y=0.35$ (red dotted line) and the mixed population
$Y_{MIXED}$ (blue dot-long dashed line). The black solid line and the
red dotted line are nearly superimposed, while the third line is
tilted with respect to the other ones, particularly for $\log P>
1.0$. This is more evident in the bottom panel \text{of the same
  figure}, where the difference between the empirical $M_k$ relation
by \cite{storm} and our theoretical relation obtained from a
\emph{pure} $Y=0.25$ (red short-dashed line) and from the mixed
population (blue dot-long dashed line) are shown.

The PWRs show the same behaviour (Fig.~\ref{mix} right panel); the
empirical $W(V,K)$ relation is by \cite{ripepi12}.  The error in the
distance determinations from the PWRs could be as high as 10\% (or
more) for $\log P$>1.5. At shorter periods the error decreases, for
example at $\log P$=1 it is about 7\%. These percentages are similar
in the different bands.

  The unexpected behaviour in
  the case of the mixed population is due to the different
  distribution of Cepheids. In fact, by
  modelling the distributions one finds: \\

\begin{description}

\item[-- at $\log P< 0.5$] Cepheids with $Y=0.25$ are brighter
    and more numerous than ones with $Y=0.35$:\\
    $N_{Y=0.25}(\log P < 0.5) = 8503$,\\
    $N_{Y=0.35}(\log P <0.5) = 0$;\\

\item[-- for $0.5\leq \log P \leq 1$] the Cepheids with
    $Y=0.25$ are brighter and less numerous than the ones with
    $Y=0.35$ ($\Delta W(V,K)$) $\sim$ 0.15 mag):\\
    $N_{Y=0.25}$(0.5$\leq$$\log P $ $\leq$1.0) = 1193,\\
    $N_{Y=0.35}$(0.5$\leq$$\log P $ $\leq$1.0) =7171; \\

\item[-- for $1.0<\log P\leq 1.5$] the number of stars for both
    $Y=0.25$ and $Y=0.35$ are much less numerous than that found at
    shorter periods: \\
    $N_{Y=0.25}(1.0<\log P \leq 1.5$) =144, \\
    $N_{Y=0.35}(1.0<\log P\leq 1.5) = 409$; \\

\item[-- for $1.5<\log P\leq 2.0$] the number of Cepheids are much
    less numerous than ones found at shorter periods:\\
    $N_{Y=0.25}(1.5<\log P\leq 2.0) = 76$,\\
    $N_{Y=0.35}(1.5<\log P\leq 2.0) = 59$.\\

\end{description}
The distributions described above show that the mixed population
considered here is composed of two major bulks of Cepheids, one at
$\log P<0.5$ ($Y=0.25$) and the second one at $0.5<\log P <1.0$
($Y=0.35$), with the latter fainter than typical Cepheids with
$Y=0.25$. This forces a simple linear fitting to provide a relation
more tilted than that from a \emph{pure} He abundance population.

In other words, at $\log P \lsim 1$ a sample fully composed by
Cepheids with a standard and a sample with enhanced He content produce
very similar PWRs. Therefore, if the samples used to calibrate the
relations contain, in addition to Cepheids with standard He abundances
($Y=0.24-0.25$), variables with a He abundance up to 0.35, these
latter stars do not affect the calibration relations significantly.  A
different behaviour is expected at longer periods, where a high
contamination of He-enhanced Cepheids (>40\%) may affect the derived
relationships.  Of course, this results require further work with
different mixture of stellar populations. Nevertheless, it provides
interesting warning against simple interpolation between pure
populations to infer indications on properties of mixed population.

\section{Comparison with Cepheids in LMC}
\label{sec:ogle}
\begin{figure*}
\center
\includegraphics[width=.68\columnwidth]{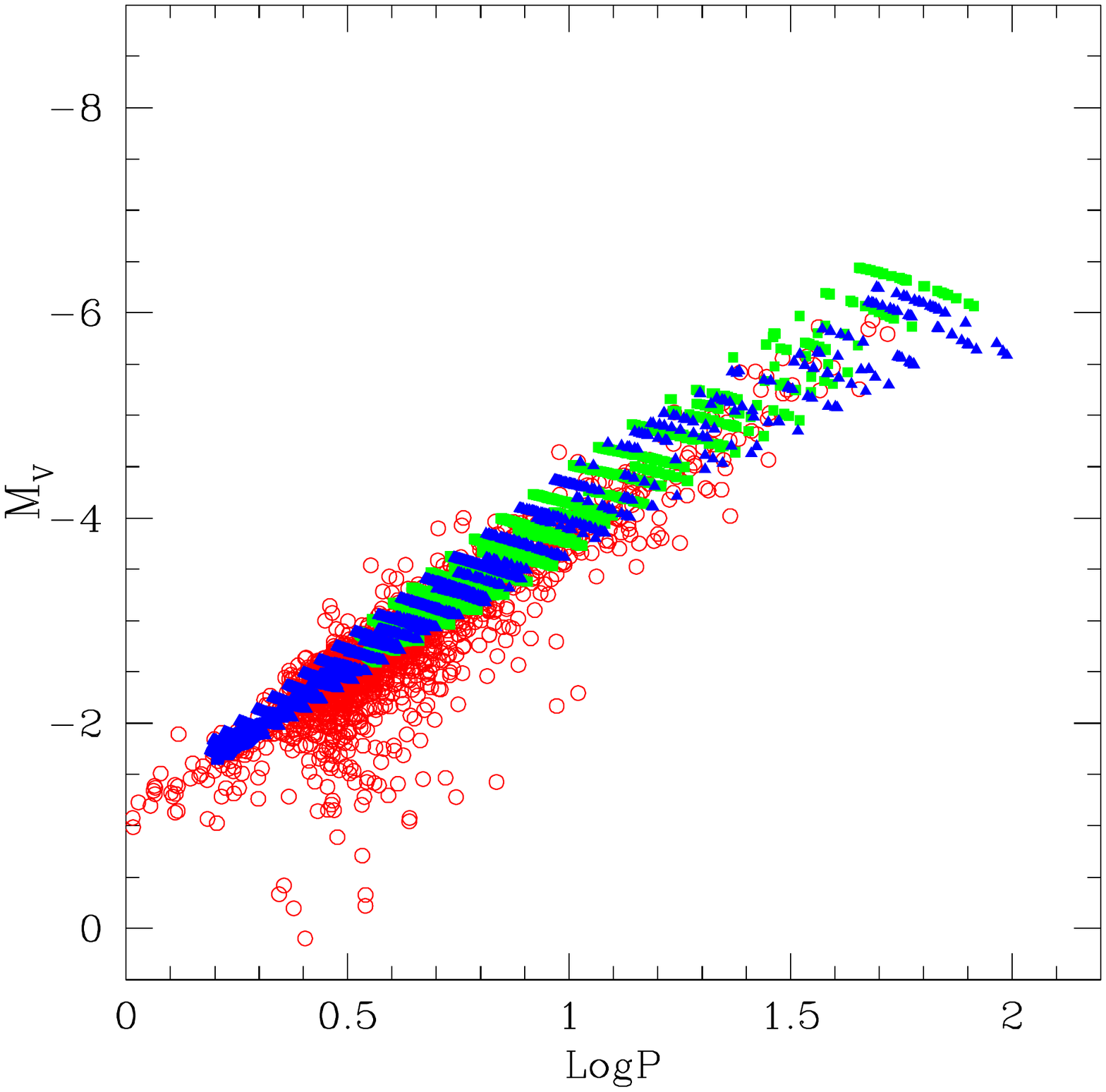}
\includegraphics[width=.68\columnwidth]{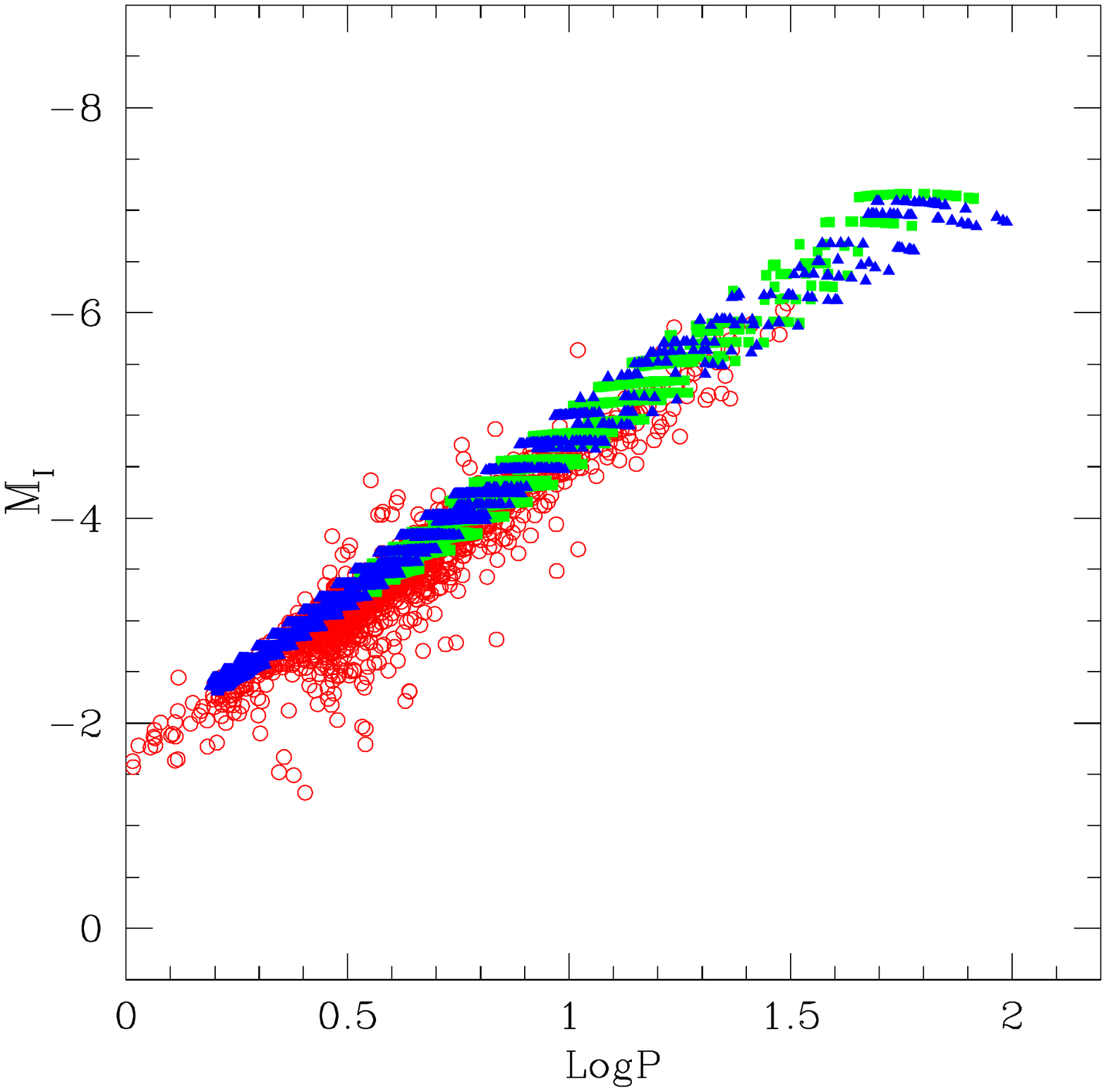}
\includegraphics[width=.68\columnwidth]{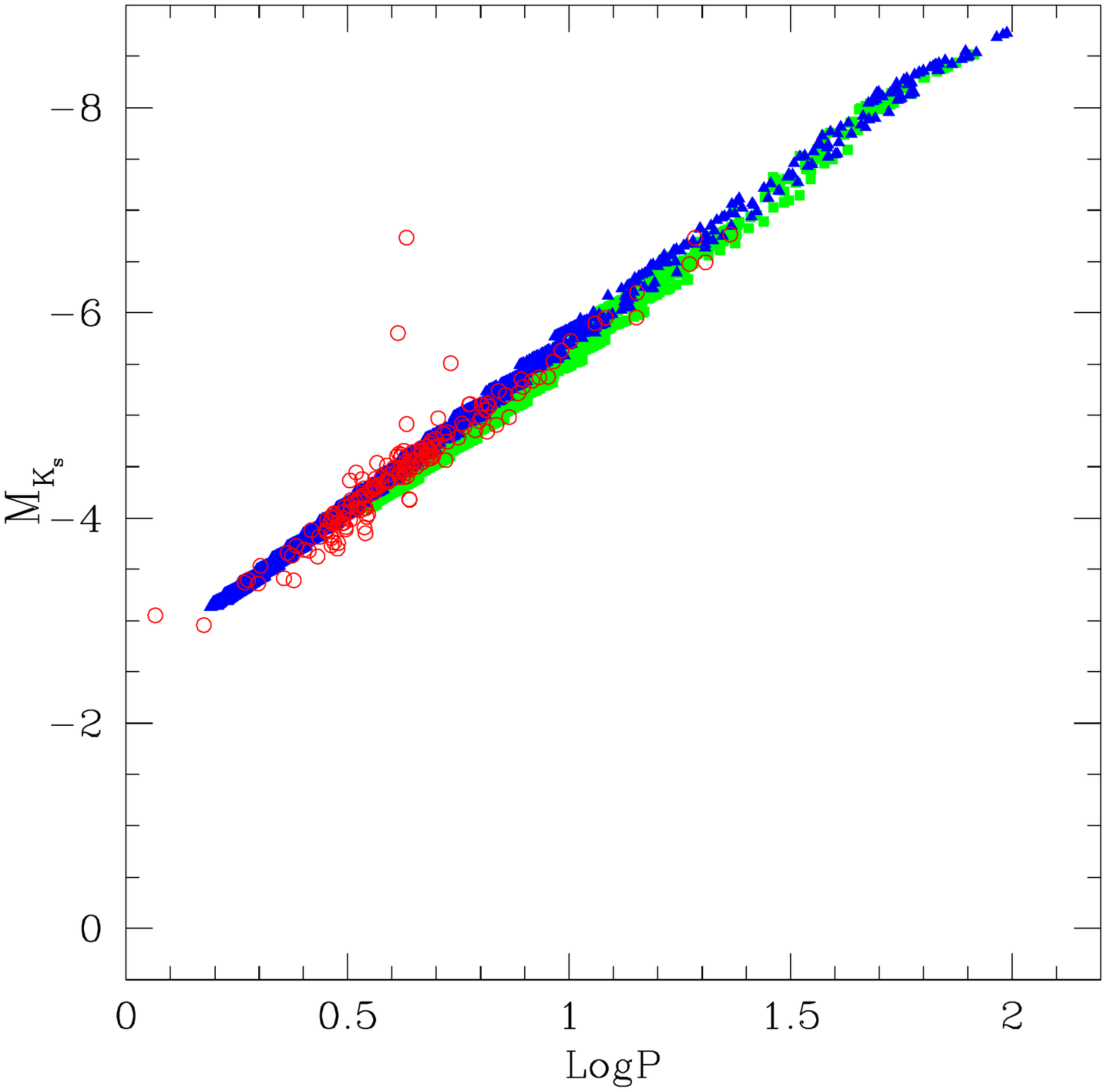}
\vspace{-1cm}
\caption{Period-Luminosity diagram in the $V$ and $I$ and $K_s$ bands
  for the classical fundamental-mode Cepheids in the LMC, taken from
  the catalogue OGLEIII ($V$ and $I$) and VMC ($K_s$) (red circles).
  The blue triangles and green squares are the simulation done with
  $Y=0.25$ and $Y=0.35$, respectively. }
\label{PI}
\end{figure*}

We used our models to search for systematic effects possibly
associated to the presence of Cepheids with higher He abundances with
respect to the primordial value in an observed sample. As a true
observational sample, we adopt the Cepheids of the OGLE III catalogue
\citep{sosz08} for the data in the V and I band, and the VMC
\citep{cioni11,ripepi12} for the Ks magnitudes. The first catalogue
includes 3361 classical Cepheids, 1848 of them pulse in the
fundamental mode. The stars are detected in the $V$ and $I$ bands, the
accuracy of the photometric calibrations is better than
0.02\,mag. Since the Cepheids are distributed along all the LMC, a
differential reddening is expected (e.g \citealt{sosz08}
\citealt{zar}).  The presence of a differential reddening affecting
the LMC Cepheids in the OGLE sample broadens the observed sequences
(see below) and the uncertainty on the apparent magnitude increases.
For example we estimated from Fig.~\ref{PI} a spread of the order of
0.2\,mag at $\log P =0.6$.  The second catalogue is described in Sec
\ref{wasenrel}.

\begin{figure}
\center
\includegraphics[trim= .1cm .1cm 1cm .1cm, clip=true,width=.9\columnwidth]{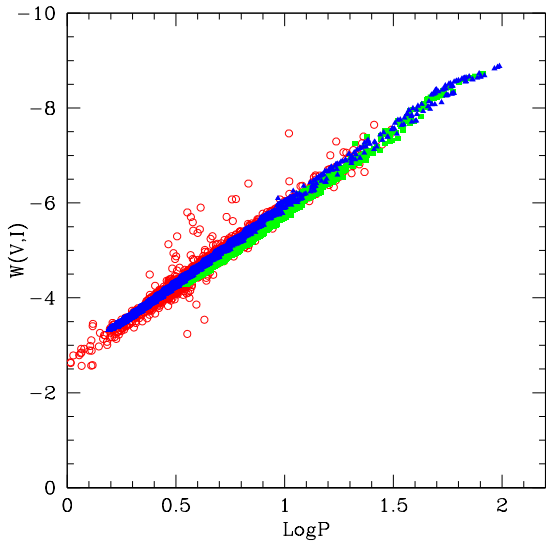}
\vspace{-1cm}
\caption{Period-W(V,I) diagram for the classical fundamental-mode
  Cepheids in the LMC, taken from the catalogue OGLEIII (red circles). Symbols are the same as in the previous figures.}
\label{P-WI}
\end{figure}

In Fig.~\ref{PI} observational data (red circles) are compared with
Cepheids predicted by models described in Sect.~\ref{sec:popmodels}
with both He abundances ($Y=0.25$, blue triangles; $Y=0.35$, green
squares), in the $V$ (left panel), $I$ (central panel) and $K_s$
(right panel) bands.

We note that a detailed simulation of LMC SFR in beyond the aim of
this paper. Differently from what assumed in our models, where a
constant SFR in the age interval 20-250 Myr is adopted, recent LMC SFR
estimations are found to vary greatly from subregion to subregion
(e.g. \citealt{rubele2012}, and reference therein), hence, we prefer
to not introduce a further not well known parameter in the present
discussion.  The predicted variables occupy the same region of the
diagram independently of the He content, and all of them overlay the
observed ones, therefore, the presence of He-enhanced stars in the
observed sample is not to be excluded.  From the comparison, it is
clear that in the period range typical of LMC Cepheids, it is
difficult to disentangle the two He abundances. We stress that the
comparison is only a by-eye analysis (including projection effect and
differential reddening effects), while the coefficients of the PLRs
and of the PWRs offer a more robust test for disentangling the two
populations. From them, it appears that the use of a He-enhanced
population gives discrepancies with what is observed, supporting the
lack of a He-rich population in the LMC, together with different
initial condition in the formation of the explored LMC regions and
globular clusters, as well as it was found for open clusters
\citep{bragaglia14}.

To investigate in more details the possible presence of He-enhanced
Cepheids, we need to study simultaneously the period, the age and the
magnitude of variables. This is because He-enhanced stars evolve more
rapidly, they are brighter and pulsate with a longer period with
respect to Cepheids of same mass but with $Y=0.25$. Note that, as we
have already shown, Cepheids with $\log P \lesssim 0.5$ cannot be
He-enhanced stars.

Our models also predict a non negligible number of Cepheids ($\sim200$
for $Y=0.25$ and 60 for $Y=0.35$) at $\log P >1.5$ and many (hundreds)
with $\log P <0.4$ not present in the OGLE III catalogue. The
long-period Cepheids are very bright ($M_I\lesssim -6.0$\,mag) and
have an age $t \lesssim 40$\,Myr, the short periods ones have $M_I <
-2.0$\,mag and $t>150$\,Myr. The lack of Cepheids observed at $\log P
> 1.5$, could be due to several reasons including a bias effect in the
observations. In the OGLE III catalogue the most bright stars ($I>13$
mag) are saturated in the $I$ band and they are not measured in the
$V$ filter, for the latter filter the saturation starts to be relevant
at periods longer than $\sim50$\,days (Fig.~\ref{PI}, right panel).
Moreover, in the OGLE III catalogue the Cepheids with $\log P < 0.4$
suffer of incompleteness because in this domain the PL sequences
overlap with various types of pulsating variables, so it is difficult
to distinguish the characteristic shapes of the light curves of
Cepheids.  In the VMC catalogue the magnitude limits in the $K_s$ band
of the Cepheids sample are $11.4 \lesssim K_s \lesssim$ 20.7 mag.

From the figure, we also note a large dispersion in magnitude in the
$V$ and $I$ bands, which is due to differential reddening as we have
already mentioned. This is also supported by the small dispersion
found in the $K_s$ band (see left panel of Fig. \ref{PI}). Further, if
we use the Wesenheit relations this spread disappears, as shown in
Fig.~\ref{P-WI} (the symbols are the same of Fig.~\ref{PI}). Again,
simulated Cepheids with both the He abundances overlap very well the
observational data.

 \section{Stochastic effects due to the number of Cepheids}
\label{sec:popeffects}

\begin{table*}
\centering
\caption{Mean standard deviations of the differences between the simulated magnitude of stars in each mini-sample and that calculated from the reference relations, representing the distance modulus $dM_X$, where X refers to the generic band.}
\label{delta}
\begin{tabular}{cccccccccc}
\hline
 Y&$\sigma (dM_B) $ & $\sigma (dM_V) $ & $\sigma (dM_I)$ & $\sigma (dM_ J)$ & $ \sigma (dM_K) $ & $\sigma (dM_{W(B,V)})$ & $\sigma (dM_{W(V, I)})$ & $\sigma (dM_{W(V,K)})$ & $\sigma (dM_{W(J,K)})$  \\
 & (mag) & (mag) & (mag) & (mag) & (mag) & (mag) & (mag) & (mag)  \\
\hline
0.25&0.14 & 0.10 & 0.07 & 0.05& 0.04& 0.04&0.04 &0.03 &0.03 \\
\hline
0.35&0.22 & 0.17 & 0.12 & 0.09 & 0.06 &0.04 &0.05 &0.05 &0.04 \\
\hline
\end{tabular}
\end{table*}


\begin{figure*}
 \center
\includegraphics[width=.9\columnwidth]{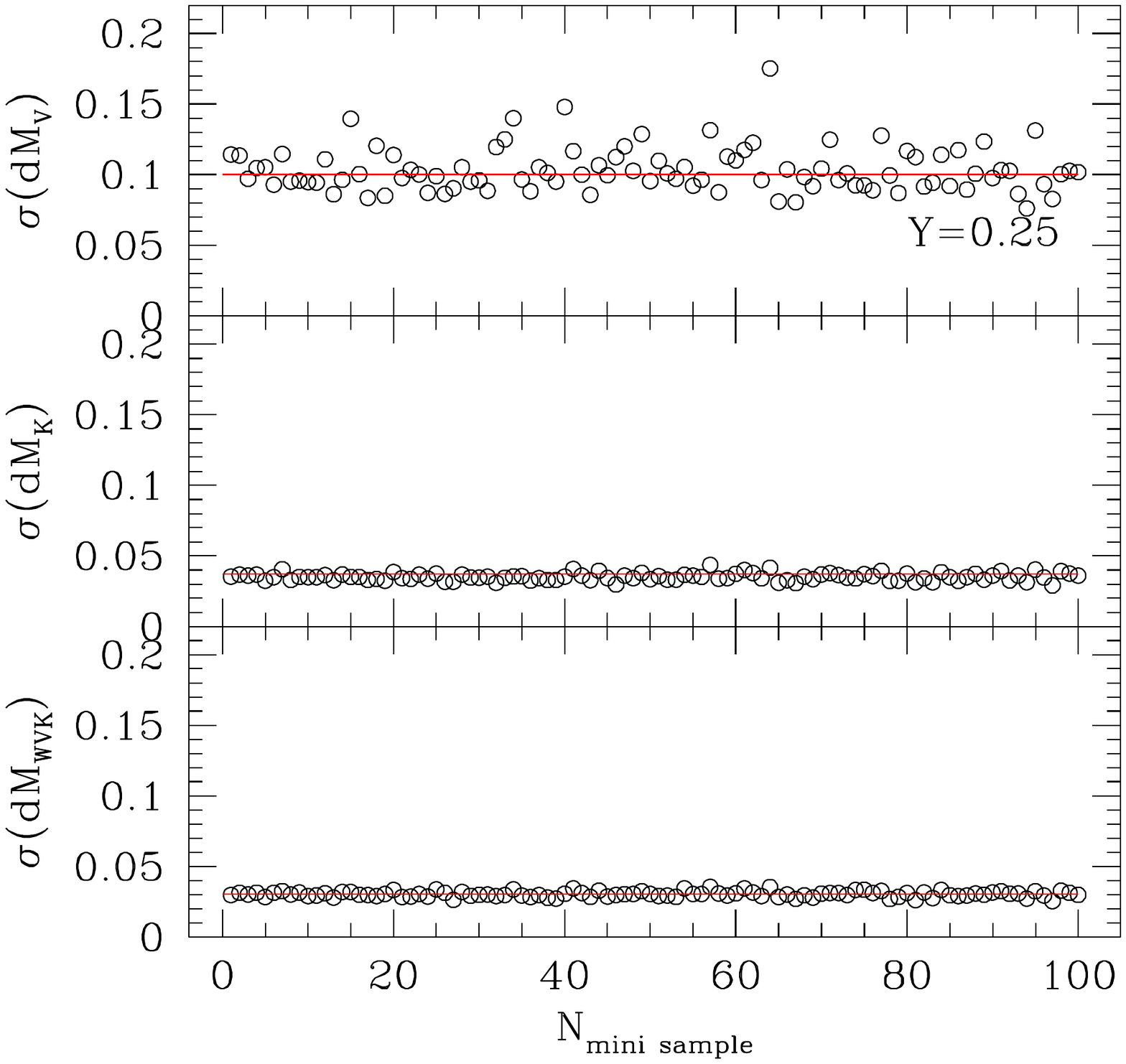}
\includegraphics[width=.9\columnwidth]{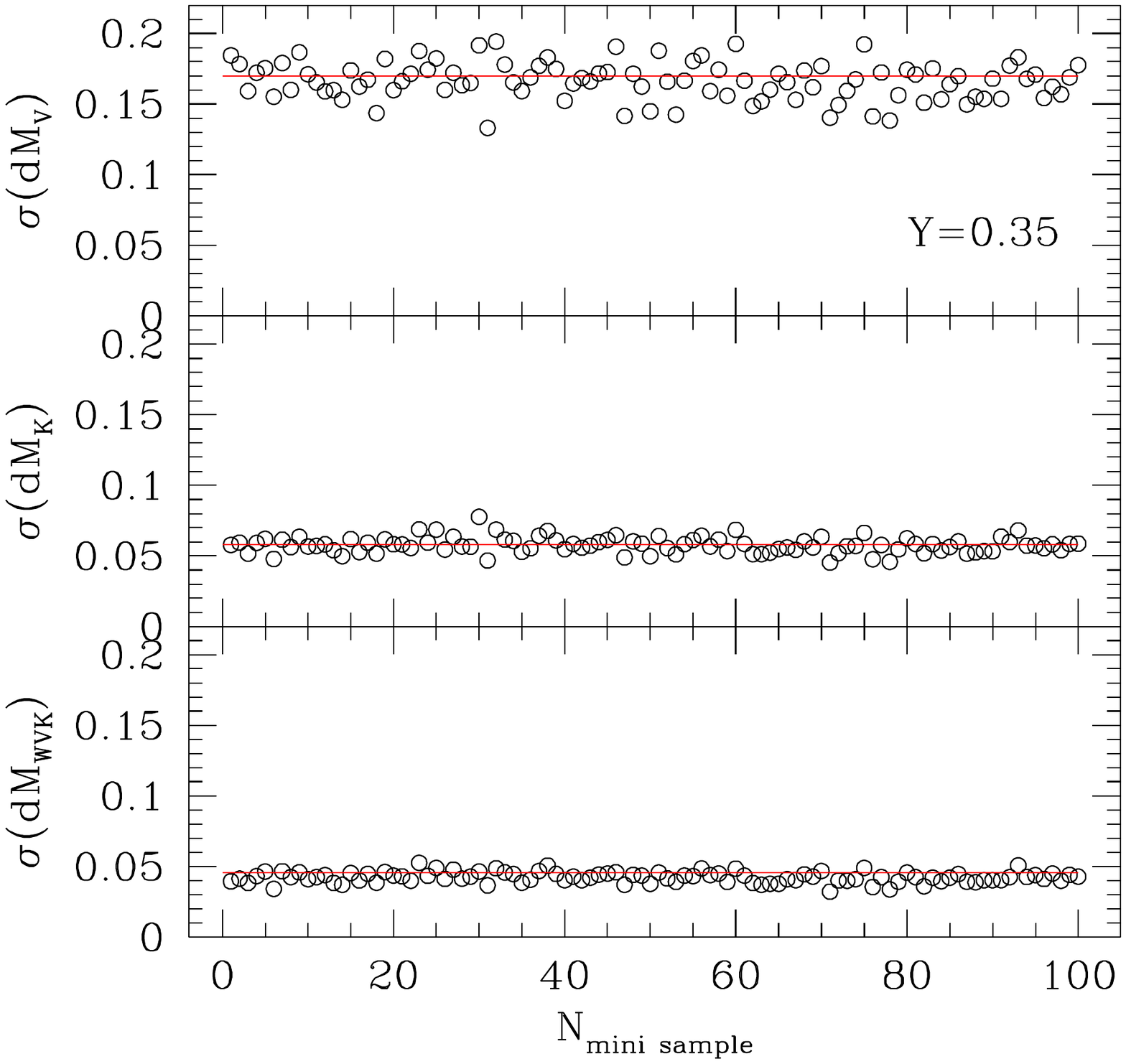}
\vspace{-1cm}
\caption{Standard deviations of the differences between the magnitude
  calculated with the reference relation and that derived from
  evolution models. This procedure is done for each mini-sample in the
  $V$ band (top panels) and $K$ band (middle panels), and for $W(V,K)$
  in the lower panels. The red horizontal lines are the mean values of
  standard deviations. Results for $Y=0.25$ ($Y=0.35$) are reported in
  the left (right) panels. }
\label{sigma}
\end{figure*}

In this section we analyse how the number of Cepheids included in the
sample affects the determination of the slope and zero point of PL and
PW relationships, that is the distance evaluation. To this purpose, by
taking advantage of the Monte Carlo procedures used to simulate
stellar populations, we used the results obtained from our sets of
models. We assembled together the Cepheids belonging to each
simulation at different ages and fixed chemical composition, ending up
with 100 independent samples of Cepheids. Each sample contains nearly
50-100 Cepheids. Hereinafter, we refer to these small samples of
Cepheids as mini-samples.

\begin{table*}
\centering
\caption{Mean, minimum and  maximum  values of the slopes and intercepts
of the $\log P$-mag and $\log P$-W relations with the respective errors (X= $a$ $\log P$ + $b$).}
\label{tabl1}
\begin{tabular}{ccccccccc}
\hline
Band & $a_{mean}$ &$\sigma_{a}$ & $b_{mean}$  & $\sigma_{b}$& $a_{min}$ &$b_{min}$ & $a_{max}$& $b_{max}$  \\
\hline
\multicolumn{9}{c}{$Z=0.008$  $Y=0.25$} \\
\hline
B & -2.634 &0.177&-0.626& 0.053&-3.040&-0.518&-2.217&-0.759\\
V & -2.859&0.125&-1.125&0.037&-3.141&-1.050&-2.570&-1.217\\
 I&-3.018&0.088&-1.730&0.026&-3.210&-1.679&-2.815&-1.787\\
 J&-3.144&0.057&-2.136&0.017&-3.267&-2.104&-3.000&-2.176\\
 K&-3.252&0.032&-2.471&0.010&-3.320&-2.450&-3.161&-2.495\\
W(B,V)&-3.556&0.058&-2.670&0.018&-3.662&-2.635&-3.436&-2.705\\
W(V,I)&-3.262&0.036&-2.662&0.011&-3.333&-2.641&-3.163&-2.689\\
W(V,K)& -3.303&0.025&-2.646&0.008&-3.351&-2.632&-3.235&-2.663\\
W(J,K)&-3.314& 0.023&-2.664&0.007&-3.357&-2.651&-3.249&-2.682\\
 \hline
\multicolumn{9}{c}{$Z=0.008$  $Y=0.35$} \\
\hline
B&-2.591&0.183&-0.799&0.123&-3.149&-0.441&-2.225&-1.016\\
V& -2.857&0.134&-1.174&0.090&-3.266&-0.911&-2.591&-1.326\\
 I&-3.049&0.098&-1.682&0.066&-3.347&-1.490&-2.854&-1.796\\
 J& -3.191&0.073&-2.025&0.049&-3.410&-1.882&-3.049&-2.107\\
 K&-3.314&0.051&-2.298&0.034&-3.458&-2.203&-3.204&-2.361\\
 W(B,V)&-3.680&0.052&-2.339&0.035&-3.807&-2.261&-3.534&-2.428\\
W(V,I)&-3.346&0.049&-2.466&0.033&-3.472&-2.382&-3.236&-2.531\\
 W(V,K)& -3.374&0.043&-2.444&0.029&-3.491&-2.357&-3.268&-2.521\\
 W(J,K)&-3.386&0.041&-2.455&0.028&-3.507&-2.366&-3.279&-2.532\\
 \hline
\end{tabular}
\end{table*}

The standard procedure to derive distance moduli  is to determine the magnitude difference
  between the individual star and the adopted reference relation and
  then average these values.  Following this procedure for each
  mini-sample, we quantify the stochastic uncertainties due to the
  small number of Cepheids of a given stellar population.  To this
  aim, we computed the difference between the magnitude of the
  Cepheids of each mini-sample with the one obtained using the
  reference PL relations for each photometric band (Table
  \ref{tablemean}).  For each mini-sample, the average of these
  differences is obviously equal to zero ($\sim$ $10^{-4}$) because we
  are dealing with absolute magnitudes. What is interesting is the
  standard deviation ($\sigma$) which provides the theoretical
  evaluation of the stochastic uncertainty.  This is shown in
  Fig. \ref{sigma}, where the $\sigma$ values in the $V$ and $K$ bands
  are plotted for all our mini-samples and for the two helium
  abundances. The average values are marked as red lines, and reported
  in Table \ref{delta}.  The uncertainties decrease from optical to
  near-infrared bands for both He abundances: for Cepheids with
  $Y=0.25$ in the $B$ and $V$ bands the $\sigma$ is as high as
  0.14\,mag and 0.10\,mag, respectively, while in the $K$ band it
  reaches 0.04\,mag. This implies that, infrared measurements remain
  quite reliable for deriving the galaxy distances even taking into
  account the stochasticity. As far as the He abundance is concerned,
  the stochastic uncertainties for $Y=0.35$ appear slightly more
  relevant than for Y=0.25. 

The procedure described above has been adopted to evaluate the
  stochastic uncertainties to derive distances with the Wesenheit
  relations and, the results are presented in Table \ref{delta}. Once
  again, the powerfulness of the Wesenheit relations is confirmed, in
  fact the values of the distance uncertainties are nearly one order
  of magnitude smaller than what found for PLRs. In particular, the
  $1-\sigma$ is of the order of 0.03 for $Y=0.25$ and $\sim$ 0.04 for
  $Y=0.35$. This can be seen in the lower panels of Fig. \ref{sigma}.

Following a different approach, the slopes of the PL and PW
  relations for each mini-sample (containing 50-100 Cepheids) are also
  computed.
Clearly, they change respect to the reference ones reported in
  Tab \ref{tablemean}, due to the smaller number of Cepheids.
Table~\ref{tabl1} lists the  mean, minimum and maximum,
values of the slopes and the intercepts for all the computed $\log
P$-L and $\log P$-W relations and their standard deviation.  As
expected, the mean values converge to the ones derived in the previous
section from a larger sample of Cepheids (see Table~\ref{tablemean}),
but here, the dispersions are larger.

Recalling that our mini-samples collect $\sim$\,50 Cepheids,
  this exercise evaluates the reliability of PLRs and PWRs when they
  are derived from small samples of variables. As a result, we find
  that samples containing few tens of Cepheids could be not adequate
  to derive the PLRs and PWRs. In fact, as representative of  worst cases,
the minimum and maximum slopes differ significantly. For instance, the
slope of the $\log P$-$M_I$ relation changes from
-3.2 to -2.8 for $Y=0.25$ and from -3.3 to -2.9 when $Y=0.35$.  Note
that the present results are consistent with the analysis presented in
\citet[][see Table 5]{carini}.  Therefore, as anticipated in
sec. \ref{sec:mixed}, the $K$ band minimizes the impact of the
statistics, both in the PLR and PWR. Variations of the He abundance do
not change this behaviour.  One should keep in mind that the magnitude
differences we have analysed can be interpreted as the
$1-\sigma$ uncertainties due to statistical effects that we
have when the distance modulus of a galaxy is derived by using a PLR
obtained from a poor sample of Cepheid stars.

\section{Summary and conclusions}
\label{sec:conclusion}

We simulated the observational properties of Cepheids belonging to
stellar populations with metallicity $Z=0.008$ and two different He
abundances, $Y=0.25$ and $Y=0.35$, in the age interval
20-250\,Myr ($Y=0.25$) and 20-150\,Myr ($Y=0.35$). With the help of
population synthesis models we investigated and quantified if and how
the presence of He-enhanced Cepheids in the observed samples could
contribute to the PLRs and PWRs and to their uncertainty.

We find that Cepheids pulsating in the fundamental mode belonging to a
stellar population with $Y=0.35$ and $\log P \lesssim
0.5$ or with $Y=0.25$ and $\log P \lesssim 0.2$ should not be
observed, because these stars, during the He-burning
phase, do not cross the instability strip.  We analysed our whole sample of
synthetic Cepheids and derived the PLRs and PWRs in different photometric
bands from the optical to near-infrared wavelengths.
The comparison between our theoretical and empirical relationships,
obtained by \cite{storm}, \cite{ripepi12} and \cite{inno} from
multi-wavelength datasets of LMC, discloses a very good agreement in
the case of synthetic Cepheids with a He abundance $Y=0.25$,
metallicity $Z=0.008$ and Salpeter IMF. 

We find that the differences in the PLRs obtained from stellar
populations having different He abundance has a negligible impact (few
percents) on the distance determination. Instead, if a mixed
population of Cepheids (composed by $\sim$\,60\% of variables with
$Y=0.25$ and $\sim$\,40\% with $Y=0.35$) is present in the observed
sample and compared to empirical PLRs and PWRs to derive distances,
the possible systematic uncertainties can be of the order of
$\sim5$-10\% at $\log P >1.5$, while at shorter periods the error is
less than 7\%, independently from the band taken into account. In the
$I$ band the uncertainty is of the order of 1\% independently from the
period. In addition, in the case of $Y=0.35$, the Wesenheit relations
are slightly tilted with respect to the empirical ones.
Therefore, in principle, when evaluating the distance of a generic
stellar population containing He-enhanced stars with the quoted
metallicity, the uncertainty due to the Helium abundance could be not
negligible, being the systematic error of the order of 5-10\% in all
bands for $\log P <1$, with the exception of the $W(V,I)$ for which
the error is as small as $\sim$\,3\%.

With the aim of investigating the presence of He-enhanced
  Cepheids in an observed sample of variables, we compared our
  simulations with Cepheids in the LMC listed in the OGLE III and the
  VMC catalogues, finding a very good agreement.  Unfortunately, there
  is not a clear separation between Cepheids having different He
  abundances, and some refinements in understanding Cepheids in LMC
  are still needed.

Finally, we studied the uncertainty introduced in the
distance estimations when one deals with a small number of Cepheids
(few tens) and derived the corresponding PL and PW relations.

Our analysis shows that the stochastic uncertainties due to
  the small number of Cepheids ($\sim$\,50) used to derive distances
  with the Wesenheit procedure is nearly negligible ($1-\sigma$
  $\lesssim$ 0.04 mag) in all bands and for both He values
  investigated here. Larger uncertainties are found for PL
  relations. Nevertheless, the $1-\sigma$ uncertainties in the $B$
  band is $\sim$ 0.14 mag for $Y=0.25$, while it becomes much smaller
  for the $K$ band ($\sim$ 0.04 mag). As far as it concerns $Y=0.35$,
  the uncertainties are slightly larger, ranging from $\sim$\,0.22 mag
  for the $B$ band up to 0.06 for the $K$ band. 
From the present analyses we conclude that, in the optical bands, the
population stochastic effect appears to be larger than the He
effect on the estimation of the PLRs and PWRs, hence on the evaluation
of the cosmological distance.  We showed that the stochastic
uncertainties are minimized in the NIR bands.

 In conclusion, the present study suggests that when information on the He
  abundance of the Cepheids in a galaxy are not available and the
  variables are compared to PLRs and PWRs obtained from local samples,
  the derived distances might be affected by non negligible systematic
  uncertainties of the order of a few percent. These uncertainties are
  more effective if the number of Cepheids observed in the given
  galaxy is small. The effect appears more severe for optical than for
  near-infrared bands. 
 Nevertheless,  we found that PLR in the K band ($\log P$ $\lesssim$ 1.5) seems to be the relation most affected by variations in the Helium content.  These results for the K
band are particularly interesting in view of future Cepheid
observations with the James Webb Space Telescope. Further fundamental
step on this issue will be also provided by the large sample of
Cepheids properties expected from the Gaia mission \citep{clementini}
and in the faraway future from the LSST survey.

\section*{Acknowledgements}
It is a pleasure to thank the referee for her/his useful suggestions and comments.
This work received partial financial support by INAF$-$PRIN2014
"\emph{EXCALIBURS: EXtragalactic distance scale CALIBration Using
  first -- Rank Standard candles"} (PI G. Clementini).
\bibliographystyle{mnras}
\bibliography{manuscriptbib}
\end{document}